\documentclass[prd,twocolumn,superscriptaddress,nofootinbib,amsmath,%
amssymb]{revtex4-2}
\usepackage{soul}
\usepackage{mathrsfs}
\usepackage{xcolor}
\usepackage{stackrel}
\usepackage{booktabs}
\usepackage{float}
\usepackage{epsfig}
\usepackage{caption}
\usepackage[font=small,justification=justified,    format=plain]{caption}
\usepackage{subcaption}
\usepackage{tikz}
\usetikzlibrary{decorations.pathmorphing}
\usepackage{hyperref} 

\begin{document}

\title{ Anisotropic Schr\"odinger black holes with hyperscaling-violation  }

\author{Alfredo Herrera-Aguilar}
\email{ aherrera@ifuap.buap.mx}
\affiliation{Instituto de F\'isica, Benem\'erita Universidad Aut\'onoma de Puebla,  Apdo. Postal J-48, C.P. 72570. Puebla, M\'exico}

\author{Jhony A. Herrera-Mendoza}
\email{ jherrera@ifuap.buap.mx}
\affiliation{Instituto de F\'isica, Benem\'erita Universidad Aut\'onoma de Puebla,  Apdo. Postal J-48, C.P. 72570. Puebla, M\'exico}

\author{Daniel F. Higuita-Borja}
\email{ dhiguita@ifuap.buap.mx}
\affiliation{Instituto de F\'isica, Benem\'erita Universidad Aut\'onoma de Puebla,  Apdo. Postal J-48, C.P. 72570. Puebla, M\'exico}

\author{Julio A. M\'endez-Zavaleta}
\email{julioamz@mpp.mpg.de} 
\affiliation{Max-Planck-Institut f\"ur Physik (Werner-Heisenberg-Institut)\\
 F\"ohringer Ring 6, 80805 Munich, Germany}

\author{Carlos Eduardo Romero-Figueroa}
\email{ cromero@ifuap.buap.mx}
\affiliation{Instituto de F\'isica, Benem\'erita Universidad Aut\'onoma de Puebla,  Apdo. Postal J-48, C.P. 72570. Puebla, M\'exico}


\begin{abstract}
We investigate novel exact solutions to an Einstein-Maxwell theory non-minimally coupled to a self-interacting dilaton-like scalar. Extending the results of \cite{Herrera-Aguilar:2020iti,Schr1}, we report three families of exact configurations over a non-relativistic Schr\"odinger background with both, arbitrary dynamical critical exponent $z$ and hyperscaling violating parameter $\theta$. Concretely, we provide  field configurations with hyperscaling violation which are asymptotically Schr\"odinger spaces. Our solutions correspond to three kinds: a zero-temperature background, a naked singularity and, more interestingly, a family of black holes. To the later, we construct the corresponding Carter-Penrose diagram with a view to understand their causal structure given the non-standard background.
We show that a non-trivial hyperscaling violation parameter $\theta$ is necessary in order to support a real non-constant dilaton field  in the configuration. We explore how the relation between the hyperscaling violation parameter and the critical dynamical exponent determine, in combination with the spacetime dimension, the kinematic aspects of the fields. We provide a thorough study of the thermodynamics including the quasi-local computation of charges, the verification of the first law and arguments concerning the stability. Lastly, we explore the effects in the thermodynamics from varying the rich parameter space. We pay special attention in comparing the qualitative behavior of the thermodynamics scalar-free solutions and the ones with a nontrivial dialton.
\end{abstract}

\maketitle


\section{Introduction\label{Sec:Intro}}

Despite the lack of a fundamental derivation, gauge/gravity correspondence, in its many different versions, has become a thought-provoking central facet in the study of theoretical high energy physics. In fact, it has recently manifested itself as a bridge to understand experimental systems from an unsuspecting  mathematical construction.  The core concept behind the prescription is the matching of the isometries of a gravity theory to that of the symmetry group of a Conformal Field Theory (CFT) defined in its boundary. Soon after the  appearance of the first Anti-de Sitter (AdS) holographic models, a potential limitation was adverted due to the intrinsic relativistic nature of gravity: a vast sector of strongly coupled systems  which happen to exhibit non-relativistic symmetry groups escaped the holographic landscape.  

In the early  works \cite{Leiva:2003kd}, various models of these kind were pointed out, with a particular interest in the case of a massless point particle over different backgrounds and potentials.
In the given examples, the classical (kinematic) symmetries contained the Galilean as well as the conformal group. After a canonical transformation, these were shown to arise at the quantum level as the symmetries of the governing Schr\"odinger equation. Soon after, this curious result motivated the challenge of constructing a geometry that explicitly realizes the Schr\"odinger group as its isometry group and accordingly dubbed the Schr\"odinger spacetime \cite{Son:2008ye}. It turns out that the resulting metric encodes a conformal relation with the AdS spacetime, hence the non-relativistic CFT (NRCFT) models  presented a sense of duality much akin to the AdS/CFT correspondence. Yet, the idea was to be developed and lacking a concrete gravity/NRCFT example. The idea was better shaped in the seminal works \cite{Balasubramanian:2008dm, Herzog:2008wg}, where a finite temperature version of the Schd\"rodinger space was introduced. Moving away from extremality allowed to establish a dictionary with low viscosity non-relativistic fluids ---effectively described by a system of strongly-coupled cold atoms. Of course, research in this direction has further proliferated,  take for instance the following recent studies: In, \cite{Wang:2013tv,Golubtsova:2020fpm},  the authors consider the dynamics of Fermi particles under unitarity. A direct computation of the boundary correlators shows that these systems also realize the Schr\"odinger group. More geometrical aspects of the gravity side were developed in \cite{Duval:2008jg,Aoki:2019bfb}. A wide introduction to the main developments can be found in \cite{Dobrev:2013kha}.

Going into the matter, the Schr\"odinger metric defined on a $D=d+3$ dimensional manifold can be  written in light-cone-light coordinates as  
 \begin{align} \label{eq:gSchr}
 ds^2_{Sch}=-b^2 r^{2z} du^2+2r^2 dudv+\frac{dr^2}{r^2}+r^2 d\vec{x}_{d}^2.
 \end{align}
 where $u,v\in [-\infty,\infty)$, $u$ is actually a null coordinate, $r\in [0,\infty)$  has a special interpretation which plays the role of an energy (holographic) dimension, and the $x^i\in (-\infty,\infty)$ cover a flat spatial  slicing. The parameter $b^2\in \mathbb{R}$ is introduced by convenience to keep track of limits of interest. Closely related to the Lifshitz spacetime \footnote{Lifshitz geometry \cite{Lifshitz} is the other pillar of non-relativistic holography. It has also an anisotropic scaling between space and time characterized by a critical exponent $z$. The main difference with the Schr\"odinger case is that, the Lifshitz group is disconnected from the AdS group except from the trivial isotropic case. }, Schr\"odinger can incorporate a critical dynamical exponent $z$ manifesting itself in accordance to the non-relativistic anisotropic scaling of space and time. In practice, this means that  \eqref{eq:gSchr}  is invariant under the transformation generated by a dilation operator
   \begin{align}\label{eq:ScaleTr}
   \mathcal{D}:\,\,  u \rightarrow \lambda^z u, \quad v \rightarrow \lambda^{2-z} v, \quad r \rightarrow \lambda^{-1} r, \quad x^i \rightarrow \lambda x^i .
   \end{align}
 In addition, as we anticipated, the Killing vectors  of  \eqref{eq:gSchr} embody and algebra closely related to the kinematic symmetry group of the Schr\"odinger equation over a flat space. The isometries' generators correspond to the momentum operators $P^{i}$, the spatial rotations $L^{ij}$, Galilean boosts $\mathcal{G}^i$, the usual time translation $H$ and  similarly the conserved mass number $N$ 
\begin{eqnarray}\label{eq:transf}
H:&  & \,\,\, u \mapsto \,\,\,u'=u+a_u;\nonumber \\
N:&  & \,\,\, v \mapsto \,\,\,v'=v+a_v;\nonumber \\
P^{i}:&  & \,\,\,x^{i} \mapsto x^{'i}=x^{i}+a^{i}, \nonumber\\
L^{ij}:&  & \,\,\,x^{i}\mapsto x^{'i}=L^{i}_{j}x^{j}, \nonumber\\
\mathcal{G}^i:&   &\,\,\,x^{i}\mapsto  x^{'i}=x^{i}-\dot{x}^{i}u, \nonumber \\
\end{eqnarray} 
where the dot notation represents total derivatives with respect to the affine parameter ---which can generically be chosen as the $u$ coordinate. The explicit form of the algebra of these generators can be found for instance in \cite{Balasubramanian:2008dm}. An inspection of it reveals that $z=2$, the pure Schr\"odinger algebra, incorporates a new generator related to the special conformal group. 

Beyond the already rich playground  provided by Schr\"odinger spaces, it is possible to broaden the scope of dual models from the NRCFT side at the cost of adding a new independent parameter. In this work, we will consider a base geometry conformally related to \eqref{eq:gSchr}. It is such that an additional constant is incorporated and denominated the hyperscaling violation exponent. That name comes from a direct connection to the comportment of the thermodynamics  boundary field theory. Hyperscaling, similar to Lifshitz scaling, refers to a measure of the dimensions carried by some quantities like the specific heat or the magnetization around a critical temperature. There are standard relations of the critical exponents to the spacetime dimension, deviations from which are named hypercaling violation, see \cite{Lesne} for a reference. 

Altogether, non-relativistic spacetimes with a hyperscaling violation exponent $\theta$ have been considered in the context of pure gravity and more concretely, in the construction of black hole and string configurations. Not surprisingly, it can be proven that such gravities require modifications to General Relativity (GR) in order to be supported. Far from being a disadvantage, these modifications can be exploited for the sake of a new viable phenomenology. To name a few examples as a motivation to our work, we have the following. In \cite{Pedraza:2018eey} a family of hyperscaling violating Lifshitz black holes supported by a framework similar to ours, an extended Einstein-Maxwell-dilaton theory, is reported.  In \cite{Dong:2012se}, the general aspects of holography involving a hyperscaling violation exponent are examined. A noteworthy result is that the exponent $\theta$ takes a major role in the presence of novel phases that violate usual entanglement entropy laws. More recently, estimations of the transport coefficients have been carried out for holographic fluid models dual to black branes with a hyperscaling violation factor \cite{Gursel:2020xou}. In these examples, it is notorious that most of the attention is devoted to Lifshitz symmetry, hence we take this window of opportunity to delve into the exploration of similar physics taking place over a Schr\"odinger spacetime. 
 
\emph{ Organization of the manuscript}.
Sec. \ref{Sec:Basics} briefly reviews some basic aspects of the geometric realization of Schr\"odinger spacetimes with the presence of a hyperscaling violating exponent. Sec. \ref{Sec:setup} presents the generalized Einstein-Maxwell-dilaton theory as the setup supporting novel hyperscaling violating Schr\"odinger backgrounds, while Sec. \ref{Sec:Solutions} is devoted to obtaining a black hole solution and its scalar free limit within this framework. Moreover, in Sec. \ref{Sec:Diagram} the corresponding  Carter-Penrose diagram is constructed to envision the global structure of this black hole configuration. In Sec. \ref{Sec:Thermo} a consistent thermodynamic analysis of our black hole is performed as well as the corresponding computation of the conserved charges. Sec. \ref{Sec:Scalarized} confronts our hairy black hole with the scalar free solution to show that they exhibit rather different thermodynamic properties. Finally, some concluding remarks are given in Sec. \ref{Sec:Remarks}.

\section{Basics on  hyperscaling violating Schr\"odinger spacetimes  \label{Sec:Basics}}

 Cast in the present coordinate system, Schr\"odinger metric \eqref{eq:gSchr} appears to be singular at the origin of the bulk coordinate. However the divergence at $r=0$ is avoidable through a suitable diffeomorphism. As a first  illustration, we compute  a few relevant curvature invariants
$R=-D(D-1)$, $R_{\mu\nu} R ^ {\mu\nu}=D(D-1)^2$ and $R_{\alpha\beta\mu\nu} R ^{\alpha\beta\mu\nu}=2D(D-1)$ which are constant and depend only   on the dimension and not on the dynamical exponent. Actually, one can show that there is a coordinate system in which the line element is regular everywhere \cite{Blau:2009gd,Lei:2013apa}, and thus a geodesically complete space.

Another interesting property of the background \eqref{eq:gSchr} is that it can be written as Kerr-Schild transformation of AdS,  as firstly studied in \cite{Duval:2008jg} and \cite{Ayon-Beato:2011rts}. Consequently, many characteristics are conferred to the geometry, among them one with great repercussion is that the Schr\"odinger spacetime can be understood as belonging to the class of exact gravitational waves over AdS. Hence, it can be shown that a general metric function replacing the coefficient of $du^2$ should obey a Siklos wave equation \cite{Duval:2008jg,Ayon-Beato:2011rts}\footnote{The Siklos operator generalizes the wave equation of flat spacetime to an AdS background. Thus, the wave profiles of the so-called AdS-waves are  a generalization of the harmonic profiles associated to pp-wave spacetimes.}.
In this sense, the constant $b^2$ takes the roll of a trivial profile  that controls the AdS limit for $b=0$. 	 

Condensed matter and other non-relativistic 
systems generically tend to introduce pertinent parameters when a more realistic physical description is needed. From the gravitational dual side, it is then useful to generalize as much as possible the AdS background to have enough freedom to match the NRQFT models. As an illustration, in condensed matter systems, a quantum critical point can be characterized by different types of critical exponents and satisfy different relationships between them. One of those interplay between parameters are the so-called \textit{hyperscaling relationships} which have the particularity of relating the scale dimensions involved in a phase transition to the spatial dimensions of the background \cite{sachdev2007quantum}. In the last years there has been a great interest in studying non-relativistic systems extending this type of property. For example, in \cite{kim2012schrodinger} it is presented a holographic realization of a non-relativistic model with an explicit hyperscaling violating symmetry together with a general dynamical exponent $ z $, focusing the study on the entanglement entropy from Schr\"odinger-type backgrounds.
The gravity side of hyperscaling violating Schr\"odinger systems can be modelled through a family of metrics conformally related to \eqref{eq:gSchr}
  \begin{align} \label{eq:gSchrHV}
 ds^2 =r^{-2\theta}\left[ -r^{2z}b^2 du^2+2r^2 dudv+\frac{dr^2}{r^2}+r^2 d{\vec{x}_{d}}^2 \right],
 \end{align}
 where $\theta$ is the \textit{hyperscaling violation exponent}\footnote{Which reduces to pure Schr\"odinger when $ \theta = 0 $, and to AdS when $\{\theta=0,b=0\}$.}.
Precisely, the hyperscaling violating factor brakes the scaling symmetry under \eqref{eq:ScaleTr}, such that, under the same transformation the line element gets an overall coefficient 
\begin{equation*}
\label{distancia}
ds^2 \rightarrow \lambda^{2\theta}ds^2.
\end{equation*}

A striking consequence due to the inclusion of the hyperscaling violating term concerns the starting discussion of this section about the global structure of the spacetime. Actually, for arbitrary values of the hyperscaling violation exponent, the singularity at $r=0$ turns out to be unavoidable. A detailed study about it can be found in \cite{Lei:2013apa}, but we put forward some basic results that will be handy in subsequent sections.  A good starting point is to write down the curvature invariants  
   \begin{align}
   R=&{}\,(D-1)(\theta-1)\left[2-(\theta-1)(D-2)\right]r^{2\theta}, \nonumber
      \\
R_{\mu\nu} R ^ {\mu\nu}=&{}\,(D-1)(\theta-1)^2\kappa_1\, r^{4\theta}, \nonumber
  \\
R_{\alpha\beta\mu\nu} R ^{\alpha\beta\mu\nu}=&\, (\theta-1)^2\kappa_2 \,r^{4\theta} ,\label{eq:Rs_HSVSchr}
\end{align}
with the constants $\kappa_i$  depending only on $\theta$ and $D$ with no information of the critical exponent $z$ and are given by 
\begin{align*}  \label{eq:kconsts}
\kappa_1=&\,(D-1)+\left[\theta(D-2)-(D-1)\right]^2,\\
 \kappa_2=&\,(D-3)^3(\theta-1)^2+2\left[3+\theta(\theta-2)\right]\\&+(D-3)\left[7+5\theta(\theta-2)\right]-2(D-3)^2(\theta-1)^2.  
\end{align*}
We note how the curvatures are either divergent or degenerate at $r=0$ or $r=\infty$ depending on the sign of $\theta$. An infinity inside the curvature invariants reveals with no doubt the presence of a singularity in the broader sense of geodesic incompleteness. Of course, the case preserving the hyperscaling ($\theta=0$) recovers our previous discussion where the singularity is just apparent.  

Another important observation with respect to the invariants \eqref{eq:Rs_HSVSchr} is that all of them vanish for the admissible and nontrivial value $\theta=1$. One might be tempted to conclude that such value removes the singularity but this idea is misleading. There is a family of geometries denominated \emph{vanishing scalar invariant spacetimes}, see \cite{Coley:2004hu} for a reference. Such metrics have the peculiarity of having zero curvature invariants at all orders despite not being maximally symmetric spaces and hence having a possible intricate global structure. Interestingly, this property for the (hyperscaling violating) Schr\"odinger space is an aftermath of the wave-like nature of its construction \cite{Duval:2008jg}.

\begin{figure}[h!]\centering
  \includegraphics[width=0.5\textwidth]{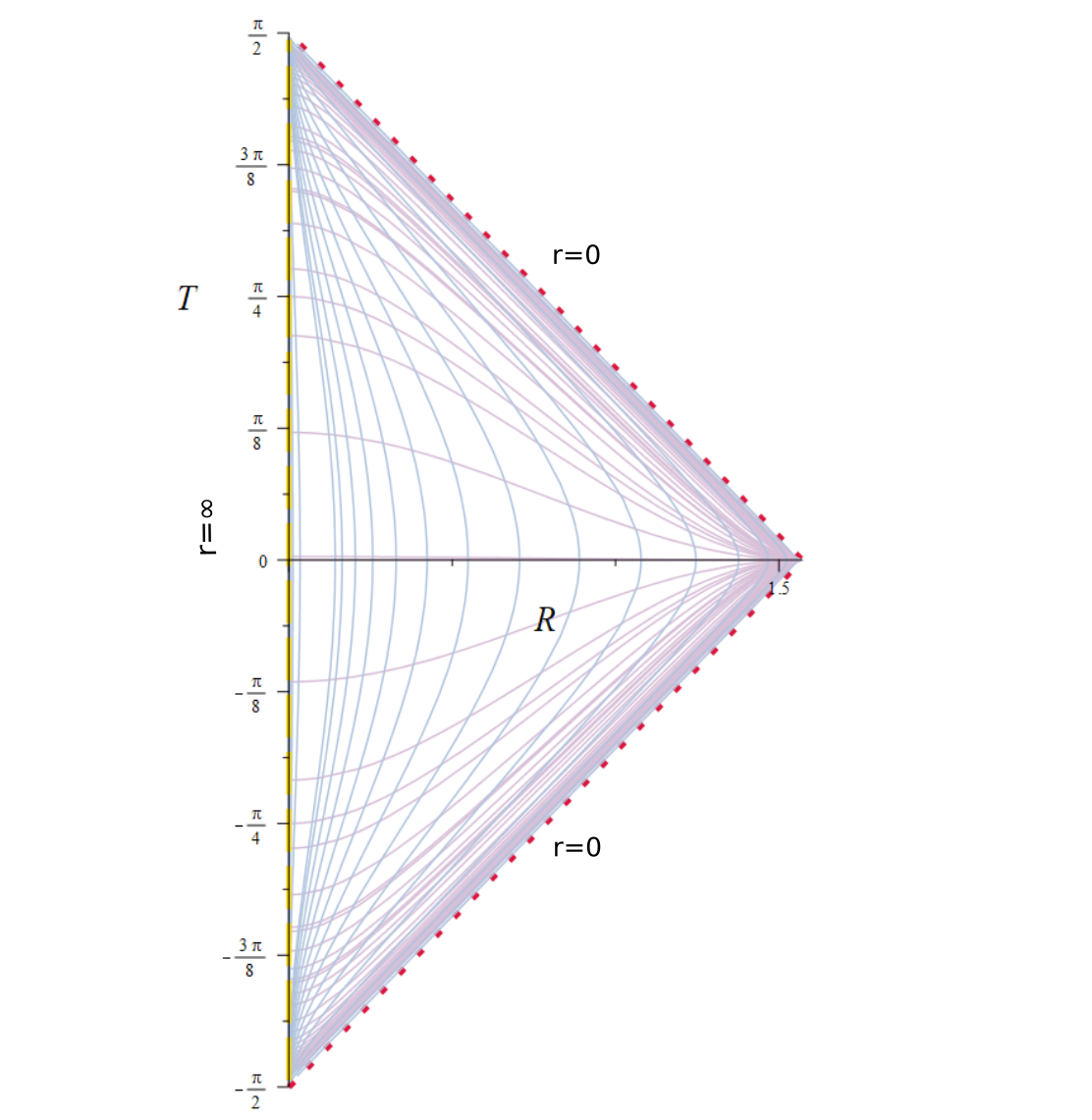} 
\caption{Carter-Penrose diagram of a generic hyperscaling violating Schr\"odinger spacetime with $\theta>0$ and $z>1$. The lilac and blue lines are the constant $u$ and $r$ null geodesics respectively represented in the compactified space. The dotted line corresponds to the spatial infinity $r=\infty$ and the yellow dashed line belongs to the singularity $r=0$ which maps to a null surface in this representation.}
\label{fig:Penrose_HSVSchr}
\end{figure}

In order to finish this argumentation, we show a simple calculation that better captures the effect of the hyperscaling violation in the global properties of \eqref{eq:gSchrHV}. Through a compactification of the spacetime similar to that of AdS or Lifshitz spacetimes \cite{Copsey:2010ya,Horowitz:2011gh}, it is possible to identify the light ray kinematics and portray the existence of a singularity. The procedure to construct a Carter-Penrose diagram Fig. \ref{fig:Penrose_HSVSchr} is rather straightforward. It comprises three steps: an identification of a tortoise coordinate, a rotation to a null plane, and a compactification via the $\arctan$ function. For simplicity we rotate back from the null coordinates obtaining the diffeomorphism 
\begin{align*}
(-\infty <u<\infty,\, &0\leq r < \infty ) \,\,\mapsto\\
&(-\pi/2< T<\pi/2,\,0< R<\pi/2).
\end{align*}
After the described coordinate transformations, the line element becomes 
\begin{align}
   ds^2=-\frac{b^2 \left(zb \sin T\cos R \right)^{2(\theta-z)/z}}{\left( \cos^2 T+\cos^2 R-1 \right)^{2(\theta+z)/z}} 
   \left( dT^2-dR^2\right).
\end{align}
 The numerator in the above expression is not problematic while the coordinates are straitened to reach the degenerate point and, other than that, there is no singular point. In the denominator,  we see that unless $\theta$ and $z$ are fine-tuned to satisfy $z+\theta=0$, there is a region where it vanishes making the singularity evident. It is clear from the Penrose diagram Fig. \ref{fig:Penrose_HSVSchr} how the geodesic can reach this singular region as pointed out in \cite{Coley:2004hu}.
 The approach here described will be the basis to examine the global properties of the black hole solutions that posses a rich structure.

\section{ Generalized Einstein-Maxwell-dilaton gravity \label{Sec:setup}}

Low energy limits of string theories provide natural modifications of General Relativity which, after exhaustive study, have become modified gravity candidates offering rich phenomenology. For instance, the axion-dilaton model which accounts for two scalar bosons serves as a strong dark matter candidate \cite{Maharana:2004qs}. We can introduce closely related theories by deforming the original structure coming from string fields. Hereafter, we will consider a generalization of the dilaton coupling to a vector field. The later can have, in principle, a gauge breaking mass term. All in all, we consider a non-minimally coupled  scalar-vector gravitational theory defined over a $D=d+3$ manifold and  encoded in the action 
\begin{align}
        S=&\frac{1}{16 \pi G}\int d^{D}x\sqrt{-g}\Bigg[R-\frac{1}{2}\partial_\mu\phi\partial^\mu\phi-V(\phi) \nonumber\\&-\frac{1}{4}\chi(\phi)F^2\ -\frac{1}{2}m^2 A^2 \Bigg],    \label{eq:actioeq:dilatonDEq}
\end{align}
where the fields have being rescaled to absorb the gravitational units,  and where $\phi$ is a real scalar field  self-interacting under the potential $V(\phi)$. There is also a vector field with field strength given by $ F_{\mu \nu}=\partial_\mu A_\nu - \partial_\nu A_\mu$ and having, in principle, a generic mass term. 
The $U(1)$ Maxwell invariance is recovered for $m^2=0$. In addition, the functions $\chi(\phi)$ stands for a coupling between the scalar and the vector, depending only on $\phi$. The dynamics of this framework is dictated by the following field equations
\begin{subequations} \label{eq:FEqs}
\begin{align}
  \mathcal{E}_{\mu\nu}:= R_{\mu \nu}-\frac{1}{2}R g_{\mu\nu}-T_{\mu\nu}=& 0,  \label{eq:EFEs} \\
  \Box \phi- \frac{1}{4}\partial_\phi\chi F^{2} -\partial_\phi V(\phi) = & 0, \label{eq:proca} 		 \\
  \nabla_\mu\Big[\chi F ^{\mu \nu}\Big]-m^2 A^{\nu} = &0,\label{eq:KG} 
\end{align}
\end{subequations}
where the energy-momentum tensor from the matter contributions has the form
\begin{align} \label{eq:Tmn}
 T_{\mu\nu}=&\frac{1}{2}\partial_\mu\phi\partial_\nu\phi-\frac{1}{4}g_{\mu\nu}{(\nabla\phi)}^2-\frac{1}{2}g_{\mu\nu}V(\phi) \\ \nonumber & +\frac{1}{2}\chi \left[F_{\alpha\mu}F^{\alpha}_{\nu}-\frac{1}{4}F^2 g_{\mu\nu}\right] \\ \nonumber &+\frac{1}{2}m^2 A_{\alpha}A_{\beta}\left(\delta^{\alpha}_{\mu}\delta^{\beta}_{\nu}- \frac{1}{2} g^{\alpha\beta} g_{\mu\nu}\right).
\end{align}

For ulterior thermodynamical applications, we target the construction of static solutions. 
It suffices then to take the following  Schr\"odinger type \textit{ansatz} with a single blackening factor $ f(r) $ and with hyperscaling violating factor
\begin{align}
\label{eq:g_ansatz}
ds^2=r^{-2\theta}\Bigg[-b^2r^{2z}fdu^2+\frac{dr^2}{r^2f}+r^2\bigg(2dudv+d\vec{x}^2\bigg)\Bigg],
\end{align}
Thus, the \emph{ansatz} (\ref{eq:g_ansatz}) is asymptotically Schrödinger as long as $ f (r) \rightarrow 1 $ near the  boundary $ r \rightarrow \infty$   when $\theta=0$. Furthermore, the dilaton and vector potential are set to inherit the isometries of the background,
\begin{equation}
\phi = \phi(r),\quad\qquad A = A_{\mu}(r) dx^{\mu}.
\end{equation}
Notice that the most general \emph{ansatz} for $A$ should include a $dv$ and $dx^i$ components when the gauge is explicitly broken ($m\neq0$).
Actually, as we explore next, the presence of the vector (Proca) mass influences the components of the vector \emph{ansatz}.

\subsection{Restoring the $U(1)$ gauge}

Before jumping in to the treatment of the complete system of equations of motion \eqref{eq:FEqs}, an observation regarding the Proca mass term, and its connection to the form of the vector potential is in order. The off-diagonal components of Einstein equations \eqref{eq:EFEs} deliver information about the first derivatives of the components of the Maxwell field. The nonzero components show up as 
\begin{subequations}\label{eq:EFEs_od}
\begin{align} 
  \mathcal{E}^{u}_{v}=  \left( A_{v}' \right)^2+ \frac{ \left(m A_{v} \right)^2}{f\chi r^{2(\theta+1)}}&=0 \\
  \mathcal{E}^{u}_{i}=    A_{u}' A_{i}' + \frac{ m^2 A_{u}A_{i} }{f\chi r^{2(\theta+1)}}&=0\\
  \mathcal{E}^{v}_{i}=    A_{v}' A_{i}' + \frac{ m^2 A_{v}A_{i} }{f\chi r^{2(\theta+1)}}&=0
\end{align}
\end{subequations}
 where the latin indices run over the $d$ transverse directions $x^{i}$
 and the prime notation denotes derivatives with respect to $r$ hereafter.
 One could in principle workout the system with full generality, however we will take advantage of an obvious simplifying assumption. Turning off the mass in \eqref{eq:EFEs_od} leads to an unique consistent and nontrivial choice
 \begin{align} \label{eq:m0_elect1}
     m^2=0 \qquad \Rightarrow \qquad A_{i}'=A_{v}' \stackrel{!}{=}0,
 \end{align}
 and with the $A_{u}$ component free. There is no straightforward condition for $A_{r}$, anyhow, when the mass is set to zero, gauge invariance is reestablished.  Hence, the $r$ component can be gauged away with the choice $A_{r}=\partial_{r}\Lambda$ for an arbitrary function $\Lambda(r)$.
 
 In short, the gauge invariant version of \eqref{eq:actioeq:dilatonDEq} has a purely electrical vector \emph{ansataz} as the most general in accordance with the isometry group
\begin{align}
A = A(r)du.
\end{align}
Of course, the previous observation is twofold. One can read condition \eqref{eq:m0_elect1} the other way around: turning off the transverse components, the Proca mass is strictly set to zero as a consequence of a purely electrical \emph{ansatz}.
With this result in hand,  a quick inspection of the Maxwell equations  reveals that it can be readily integrated once in terms of the arbitrary metric  and coupling functions
\begin{equation}
{ A}^\prime(r)=\frac{q}{\sqrt{f}\chi r^{ (D-3)(1-\theta)+\theta+1}}.\label{eq:Maxwell1st}
\end{equation}
Here $q $ is an arbitrary integration constant that is associated with the total electromagnetic $U(1)$ charge.

\subsection{Zero-temperature configuration: bare hyperscaling violating Sch\"rodinger background}

From now on, we take into consideration the result of the previous section. The vector field will play solely the roll of a Maxwell potential with strictly zero mass. In due course,  a relevant and functionally independent set of field equations reads
\begin{subequations} \label{eq:IndepSys}
\begin{align}
\mathcal{E}^v_u&=0  \nonumber \\
{}&= r^2f^{\prime\prime}-\big [ \theta(D-2)-5(z-1)-D)\big]rf^\prime \nonumber \\& +\frac{1}{2}{\frac{r^2f^{\prime2}}{f}}- 2(z-1)\big[\theta(D-2)-2z-D+3\big]f \nonumber\\&-\frac{q^2r^{2\left[(D-3)(\theta-1)-z\right]}}{b^2f\chi},\label{eq:NLEuler}\\
\mathcal{E}^r_{r}-\mathcal{E}^u_{u}&=0=  
(\phi^\prime)^2-\frac{(D-2)(\theta-1)}{r^2}\bigg[\frac{rf^\prime}{f}+2\theta\bigg],\label{eq:dilatonDEq}\\
\mathcal{E}^r_{r}+\mathcal{E}^u_{u}&=0=  
V-\frac{(D-2)(\theta-1)}{2}r^{2\theta}\Big\{rf^\prime \nonumber\\  &-2\left[\theta(D-2)-D+5)\right]f\Big\}.
 \label{eq:PotAEq}
\end{align} 
\end{subequations}
It can be indeed verified that, due to the conservation of the energy-momentum tensor (\ref{eq:Tmn}), the scalar Klein-Gordon equation provides no new information, it is functionally dependent on the above system.   

It is relevant to observe that we have an under-determined non-linear system \eqref{eq:IndepSys}, with the freedom to fix one of the unknowns conveniently to our task of finding regular solutions. The most evident choice is to study the case with no blackening function $f(r)=1$, i.e, a zero temperature solution. In this situation, we can explore the sole effect of the hyperscale violation enforced on the Schr\"odinger space. For such a choice, the system is consistent and yields the following  results: a logarithmic dilaton profile
\begin{align} \label{eq:phiT0}
\phi(r)=\sqrt{2\theta(\theta-1)(D-2)}\ln(r)+\phi_{0}
\end{align} 
which is subject to a Liouville self-interaction potential of exponential kind, similar to the ones studied in \cite{Charmousis:2001nq} 
\begin{equation}
    V(\phi)=V_0e^{\lambda_0(\phi-\phi_0)}, \label{eq:ZTpot}
\end{equation}
 and is coupled with the Maxwell field through a dilatonic coupling
\begin{equation}
  \chi(\phi)=\chi_0e^{\lambda_1(\phi-\phi_0)}.\label{eq:ZT_NMC}
\end{equation} 
In these expressions  we have introduced constants given in terms of the metric parameters and the electric charge
 \begin{subequations}
\begin{align}
V_0=&-\left[\theta(D-2)-(D-1)\right](D-2)(\theta-1),\label{q22}
\\
\chi_0=&\frac{q^2}{2(z-1)\left[D-3+2z-\theta(D-2)\right]b^2}\label{q33}
\\
\lambda_0=&\sqrt{\frac{2\theta}{(D-2)(\theta-1)}},\quad\lambda_1={\frac{2[(D-3)(\theta-1)-z]}{\sqrt{2\theta(D-2)(\theta-1)}}}.
\end{align}
\end{subequations}
We have all the information necessary to fully integrate the Maxwell equation (\ref{eq:Maxwell1st}) under the condition $f=1$. For this aim, one must inspect two separate branches of solutions in accordance to a restriction in the parameter space. Concretely,
\begin{equation}
A(r)=\frac{2b^2(z-1)}{q}r^{\left[D-3+2z-\theta(D-2)\right]}+A_0,
\end{equation}
 with $A_0$ an integration constant which can be turned off without loss of generality.
Due to the lack of a horizon as advertised in Sec.~\ref{Sec:Basics}, this configurations presents a trivial thermodynamics. The temperature is constantly vanishing as well as the entropy. However, we can still infer some noteworthy effects due to dynamics scaling and hyperscaling violation exponents. First, notice the bare Schr\"odinger limit $\theta \rightarrow 0$. The most obvious aftermath is that the dilaton field is set to be constant and therefore trivial. Acting carefully the limit, one also has that the self-interaction potential becomes an effective cosmological constant and the  non-minimal coupling a constant rescaling the net charge. We spot the value $\theta=1$ as another singular point. Evaluated there, the scalar also trivializes. Anyhow, in that case the potential and  the dilatonic coupling are also rendered constants not dynamically determined by the field equations.  
There are other regions in the $\theta -z$ plane that fundamentally changes the nature either of the fields or the free functions. We summarize the different results Table \ref{tab:ZeroTemp}.
\begin{table}
\begin{tabular}{@{} *4l @{}}    \toprule
\emph{Condition} & \emph{dilaton}  ($\phi$) & \emph{Coupling} ($\chi$) &  \emph{Potential} ($V$) \\\midrule
$\theta=0$   & const  &   const  & const$<0$    \\ 
 $\theta=1$  & const &  const & 0  \\ 
 $\theta=\frac{D-1}{D-2}$ & log & const $\sim (z-1) $ & 0 \\
$ \theta=\frac{D+2z-3}{D-2}$ & log & arbitrary &  const $\sim (z-1) $\\
$\theta=\frac{D+z-3}{D-2} $ & log   & const  & exp \\
 \bottomrule
 \hline
\end{tabular}
 \caption{\label{tab:ZeroTemp} We show the special values and regions in the parameter space  where the  field equations display a different branch of solution. Abbreviation are as follows: log $=$ logarithmic, const $=$ constant, exp $=$ exponential. }
\end{table}

Notice that in the last $\theta-z$  relation of  Table \ref{tab:ZeroTemp},  there is no need  for an explicit interaction between the scalar and vector sectors. The non-minimal function becomes just a normalization of the vector field and thus, the scalar field, minimally coupled to Einstein-Maxwell suffices to support the Schr\"odinger spacetime with that particular hyperscaling violating factor.

\section{Exact static black hole solution
\label{Sec:Solutions}}

Configurations with horizon are the central object of study of this work. Unlike the solution already presented, a black hole will be prone to a well-defined thermodynamics and thus posing results towards the Schr\"odinger/NRQFT correspondence. This fact was already addressed in \cite{Herzog:2008wg}, where the search for asymptotically Schr\"odinger black holes was initiated. 
In this section we take advantage again of the freedom in the equations of motion. With this aim, the best strategy is to take advantage of the non-minimal coupling function. We will investigate two feasible instances in which the non-linear equation \eqref{eq:NLEuler} becomes solvable for $f$.  A first solution, a family of hyperscaling violating black holes is found by means of the following fixing of the coupling
\begin{equation} 
    \chi(\phi)=\frac{2 (q/b)^ 2 r^{2\left[(D-3)(\theta-1)-z\right]}}{
    \big(rf^{\prime}\big)^2-4(z-1)[\theta(D-2)-D-2z+3]f^2 }\label{eq:chiBH0},  
\end{equation}
The other notorious choice, elaborated in Appendix \ref{sing}, leads to a less physically relevant naked singularity.

 Plugging the form of the coupling function (\ref{eq:chiBH0}) into \eqref{eq:dilatonDEq} linearizes the differential equation bringing it to a second-order Euler equation
\begin{equation}
  r^2f^{\prime\prime}-\bigg[\theta(D-2)-5(z-1)-D\bigg]rf^\prime=0 ,\label{hn} 
\end{equation}
 The most general solution can be readily tackled, which, after a proper fixing  of the integration constants is 
\begin{equation}
    f(r)=1- \bigg(\frac{r_h}{r}\bigg)^{\beta},\label{eq:solf_BH}
\end{equation} 
where we have introduced 
\begin{align} \label{eq:beta}
\beta:=  2\theta -D(\theta-1)+5z-6>0
\end{align}
as a strictly positive  constant in order to ensure the desired asymptotic Schr\"odinger geometry.
For the same reason, one of the integration constants was rigidly chosen as $1$. The remaining constant was pinpointed as the position of the event horizon $r=r_{h}$. The solution (\ref{eq:solf_BH}) has all free parameters characterizing the dual non-relativistic theory. Namely, it has information of  the unfixed critical dynamical exponent $z$, the hyperscaling violating exponent $\theta$ and an arbitrary number of dimensions $D-1$ over the boundary.  

We move on to determining the scalar field compatible with this geometry. With $f$ given as in (\ref{eq:solf_BH}), the dilaton can be explicitly integrated from (\ref{eq:dilatonDEq}). The outcome for the dilaton's profile is
\begin{align} \label{eq:phiBH}
\phi(\rho)=& \frac{ 2\sqrt{(\theta-1)(D-2)} }{\beta} \Bigg[ \sqrt{\alpha} \text{arcsinh}\left(  \sqrt{ \frac{\alpha}{\beta} \left(1-\rho^\beta\right) } \right) \nonumber \\ &-\sqrt{2\theta}\text{arctanh}\left( \sqrt{ \frac{ 2\theta(1-\rho^\beta)}{ 2\theta-\alpha \rho^\beta  }}\right) \Bigg]+\phi_0
\end{align}
where we have introduced the short-hard notation
\begin{align}
\rho=\frac{r_{h}}{r}, \qquad\qquad \alpha= 2\theta-\beta,
\end{align}
 and $\phi_0$ is an arbitrary integration constant that later takes an interesting character in the thermodynamics.
Our result \eqref{eq:phiBH} turns out quite rich with different aspects to be discussed.
For instance, we start by noting  that $\phi$ has a finite and real value on the horizon $r=r_h$ ($\rho=1$), independently of the values of the hyperscaling violation and critical exponent. At spatial infinity, which is to say, at the boundary of the spacetime $r\rightarrow\infty$ ($\rho\rightarrow0$), the dilaton has two possible behaviors. For an arbitrary $\theta$, the $\text{arctanh}$ contribution is problematic, since it is divergent when the argument goes to $1$. However, one can remedy that problem when there is no hyperscaling violation, i.e, when $\theta=0$.  
 A more detailed inspection reveals that the scalar field can actually acquire real values in less obvious regions of the parameter space $\theta-z$. Depending on what region along the holographic coordinate the scalar is measured, its dynamics is well posed for different combinations of the critical and hyperscaling violation exponents. The identification of such disconnected regions --- in the parameter space--- is tricky and  we summarize it in table \ref{tab:my_label3} supplemented with further discussion below.
\begin{table}[h]
\begin{tabular}{@{} *3l @{}}    \toprule
\emph{ $\theta$ range} & $z$ range & \emph{spacetime region}  \\ \midrule
$1<\theta< D+5z-6$ & $z<\frac{6}{5}$ &  $r>r_h$     \\ 
 $1<\theta< \frac{D+5z-6}{D}$ & free &  $r>r_h$  \\ 
$\theta=\frac{D+5z-6}{D}<0$ & $z>\frac{6}{5}>0 $  &  $r> r_h$ \\
$\theta=\frac{D+5z-6}{D}<0 $ & $ \frac{6-D}{5}<z<\frac{6}{5} \quad$  &  $r<r_h$\\
$\frac{ D+5z-6}{D}<\theta<\frac{ D+5z-6}{D-2} $ & free  & $r<r_h$   \\
$0<\theta<1 $ & $ z>\frac{6-D}{5} $ & $0<r<\left(\frac{2\theta}{\alpha}\right)^{-\frac{1}{\beta}}r_h$\\
   &   \\
 \bottomrule
 \hline
\end{tabular}
 \caption{\label{tab:my_label3} Different regions of spacetime, determined by a range on $r$, where the scalar field is real and thus physically sensible depending on relations between the $(\theta,z,D)$ parameters. }
\end{table} 

In principle,  we consider the dynamics of the scalar field  taking effect in the spacetime sector where the blackening factor (\ref{eq:solf_BH}) preserves the metric signature (\ref{eq:g_ansatz}).
In that sense, our solution describes only the exterior of the black hole, before crossing the event horizon.
Notwithstanding, the profile function exhibits some allowed parameter ranges in which it is also real in the interior of the horizon. Given that it is a scalar function, the restrictions where $r<r_h$ in  Tab. \ref{tab:my_label3} should be invariant and also admissible given that proper geometry is  also constructed.
As a supplementary tool for visualizing the regions of validity provided in Tab.\ref{tab:my_label3}, we put up two density plots displaying the numeric value of the dilaton's profile as a function of the 2-dimensional parameter spacer $(\theta,z)$. This in order to get a better idea on how the hyperscaling violation and the dynamical scaling parameter collude in disconnected zones to generate a real scalar.

\begin{figure}
     \centering
     \begin{subfigure}[b]{0.22\textwidth}
         \centering
         \includegraphics[width=\textwidth]{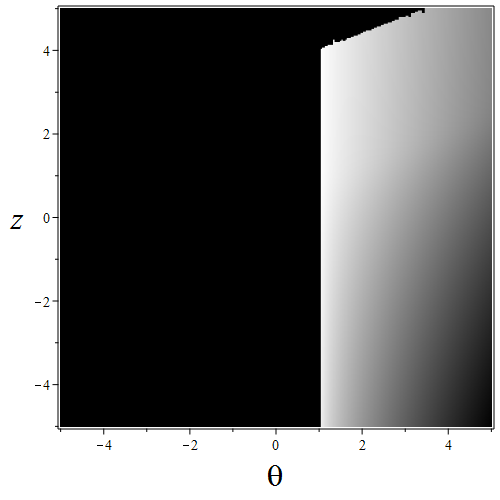}
         \caption{$D=4$}
         \label{fig:zt4de}
     \end{subfigure}
     \hfill
     \begin{subfigure}[b]{0.22\textwidth}
         \centering
         \includegraphics[width=\textwidth]{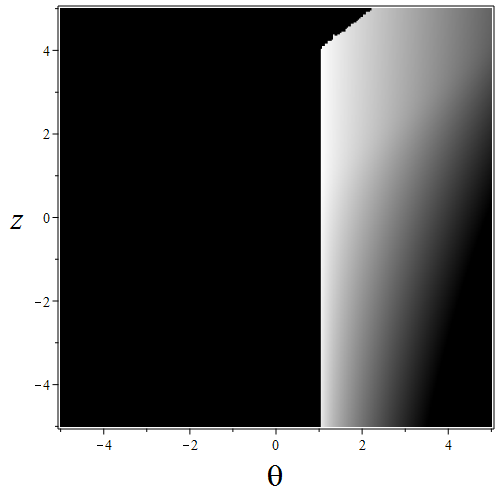}
         \caption{$D=6$}
         \label{fig:zt6de}
     \end{subfigure}
     \hfill
        \caption{A density plot of the numerical values of the scalar field's profile in the outside of the horizon (for an exemplifying fixed $r>r_h$) is displayed. The completely black patches depict regions in the parameter space where the field takes complex or unbounded (divergent) values. We show the $D=4,6$ cases from left to right for illustration.}
        \label{fig:zt-ext}
\end{figure}
\begin{figure}
     \centering
     \begin{subfigure}[b]{0.22\textwidth}
         \centering
         \includegraphics[width=\textwidth]{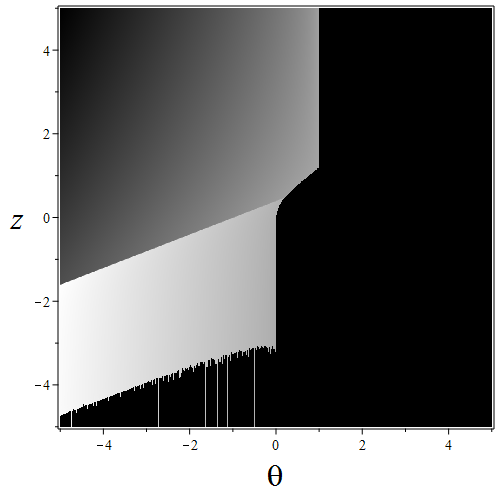}
         \caption{$D=4$}
         \label{fig:zt4de}
     \end{subfigure}
     \hfill
     \begin{subfigure}[b]{0.22\textwidth}
         \centering
         \includegraphics[width=\textwidth]{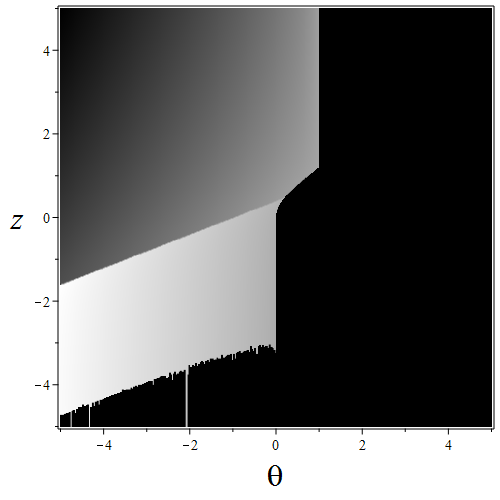}
         \caption{$D=6$}
         \label{fig:zt6de}
     \end{subfigure}
     \hfill
        \caption{A density plot of the numerical values of the scalar field's profile in the inside of the horizon (for an exemplifying fixed $r<r_h$) is displayed. The completely black patches depict regions in the parameter space where the field takes complex or unbounded (divergent) values. We show the $D=4,6$ cases from left to right for illustration.}
        \label{fig:zt-int}
\end{figure}

 The whole gravitational configuration is not yet complete.  The following is to determine its self-interaction potential as well as the the non-minimal coupling function. By inserting (\ref{eq:solf_BH}) into (\ref{eq:chiBH0}) the coupling function is implicitly determined, which for our solution takes the following form
\begin{equation}
  \chi(\rho)=\frac{2(q/b)^2 (r_h/\rho)^{2\left[(D-3)(\theta-1)-z\right]}}{\delta_1\left(1-2\rho^{\beta} \right)+\delta_2 \rho^{2\beta}}\label{ eq:chi_BH},
\end{equation}
where we have defined
\begin{subequations}
\begin{align}
    \delta_1=&4(z-1)\big[2(z+\theta)-D(\theta-1)-3\big], 
\\
    \delta_2=&\delta_1+\left[6-5z+\theta(D-2)-D\right]^2.
\end{align}
\end{subequations}
In a similar fashion, evaluating (\ref{eq:solf_BH}) in (\ref{eq:PotAEq}) grants the explicit form of the self-interaction potential
 \begin{equation}
V(\rho)=\frac{1}{2}(D-2)(\theta-1)\left(\frac{r_h}{\rho}\right)^{2\theta}\left( \delta_3+\delta_4\rho^{\beta}\right),\label{eq:Vr_BH}
\end{equation}
with the following constants definition
 \begin{align*}
     \delta_3=&-2\left[\theta(D-2)-D+1\right],\\
      \delta_4=& \,5z+\theta(D-2)-D-4.
 \end{align*}
 One must observe that both the self-interaction potential and the non-minimal coupling function are reduced to a finite constant when evaluated at the event horizon. Finally, to determine the vector potential, we insert the metric function (\ref{eq:solf_BH}) and the coupling function (\ref{ eq:chi_BH}) into the Maxwell's first integral (\ref{eq:Maxwell1st}) and fully integrate the vector field with respect to $ r $, obtaining the following solution
\begin{align}
  A(\rho)=&\frac{b^2}{q} \left(\frac{r_h}{\rho}\right)^{2z-\theta(D-2)+D-3}\sqrt{1-\rho^{\beta}}\Big[2(z-1) \nonumber \\&-\left[\theta(D-2)-3z-D+4)\right]\rho^{\beta}\Big].\label{ eq:Maxwell_BH} 
\end{align}
 Therefore, the complete solution is given by the gravitational field (\ref{eq:g_ansatz}) with a blackening factor (\ref{eq:solf_BH}). A real scalar field (\ref{eq:phiBH}) with  a self-interaction potential (\ref{eq:Vr_BH}) and non-minimal coupling (\ref{ eq:chi_BH}) with an electromagnetic field generated by the vector potential (\ref{ eq:Maxwell_BH}). The novel configuration here derived is indeed a black hole spacetime. We demonstrate this claim by exploring the global properties of the metric in Sec. \ref{Sec:Diagram}.

\subsection{The scalar free solution $\theta=1$ \label{Sec:ScalarFree}}

Recall the original system of field equations \eqref{eq:IndepSys} for the given \emph{ansatz}. Paying special attention to equations \eqref{eq:PotAEq} and \eqref{eq:dilatonDEq}, it is patent that $\theta=1$ is a singular value requiring separate treatment. The most noticeable consequence of such value is that the dynamics of the scalar sector trivializes, this is: the dilaton has a constant as a solution, the self-interaction potential strictly vanishes ---not even with a cosmological constant term--- and, consequently, the non-minimal coupling function becomes a constant too.  The last statement is transcendental according to Eq.~\eqref{eq:NLEuler} and the strategy described above to get rid of the non-linear part and facilitate its integration. More explicitly,  with $\chi=const$ it is not possible to pick it in a way that, together with the accompanying coefficients, cancels the non-linearity (and possibly the zeroth order term)\footnote{There is an unique special case in which the nonlinear term can be removed with $\chi=const$ for $f\sim r^{-z}$ and a particular dynamical exponent $z=0,1$ but it is of no interest for black hole configurations.}. 
Hence, the differential equation governing the blackening function in this particular case is
\begin{align} \label{eq:feq_full}
    0=&\,r^2f^{\prime\prime}+\big(5z-3\big)rf^\prime \nonumber  +\frac{1}{2}{\frac{r^2f^{\prime2}}{f}}\\&+ 2(z-1)\big( 2z-1\big)f  -\frac{q^2r^{-2z}}{b^2f},
\end{align}
setting $\chi=1$ without loss of generality. It is important to remark that  naively taking the limit $\theta=1$ in the fields of the black hole configuration, shows the first two points here advertised: the vanishing of the scalar and the potential. However, the metric function $f$ has no apparent alteration beyond restricting the power $\beta$ involved therein. Equation \eqref{eq:feq_full} suggests that actually, the solution of $f$ has a much more involved behaviour than \eqref{eq:solf_BH} and it is worth exploring separately.

One naive attempt test is to evaluate the blackening function \eqref{eq:solf_BH} with $\theta=1\,\Rightarrow \, \beta=5z-4 $. The result, a bit lengthy to be worth writing down, is not zero and thus $f=1-(r_h/r)^\beta$ is not a solution of \eqref{eq:feq_full}. The only possible reconciling way out is to turn off the constants $b=0$ and $q=0$, which results in a huge trivialization of the configuration. Actually, by doing so, the Maxwell field vanishes and the metric becomes diffeomorphic to AdS.

There is no general strategy known to the authors to tackle the solution of \eqref{eq:feq_full}. One could try to numerically integrate with the appropriate boundary conditions or find a near-horizon solution. Anyhow, for our purposes it is enough to make clear that there exist a different branch of solutions, say $\hat{f}(r;\theta=1)$  that has the characteristic 
\begin{align}
    \hat{f}(r) \neq \lim_{\theta\rightarrow 1} f(r),
\end{align}
understanding $f$ as in the black hole solution with scalar field. What this says is that there is no smooth limit in the parameter space that leads our hairy solution to another scalar-free black hole of the theory,  in contrast to the asymptotically Lifshitz case in \cite{Herrera-Aguilar:2020iti}. Therefore, it is not possible to study (spontaneous) scalarization within the present solution.

\section{Global structure of the black hole
\label{Sec:Diagram}}

\subsection{Sketch of the Carter-Penrose diagram for a generic parameter configuration }
\begin{figure}[h!]\centering
  \includegraphics[width=0.5\textwidth]{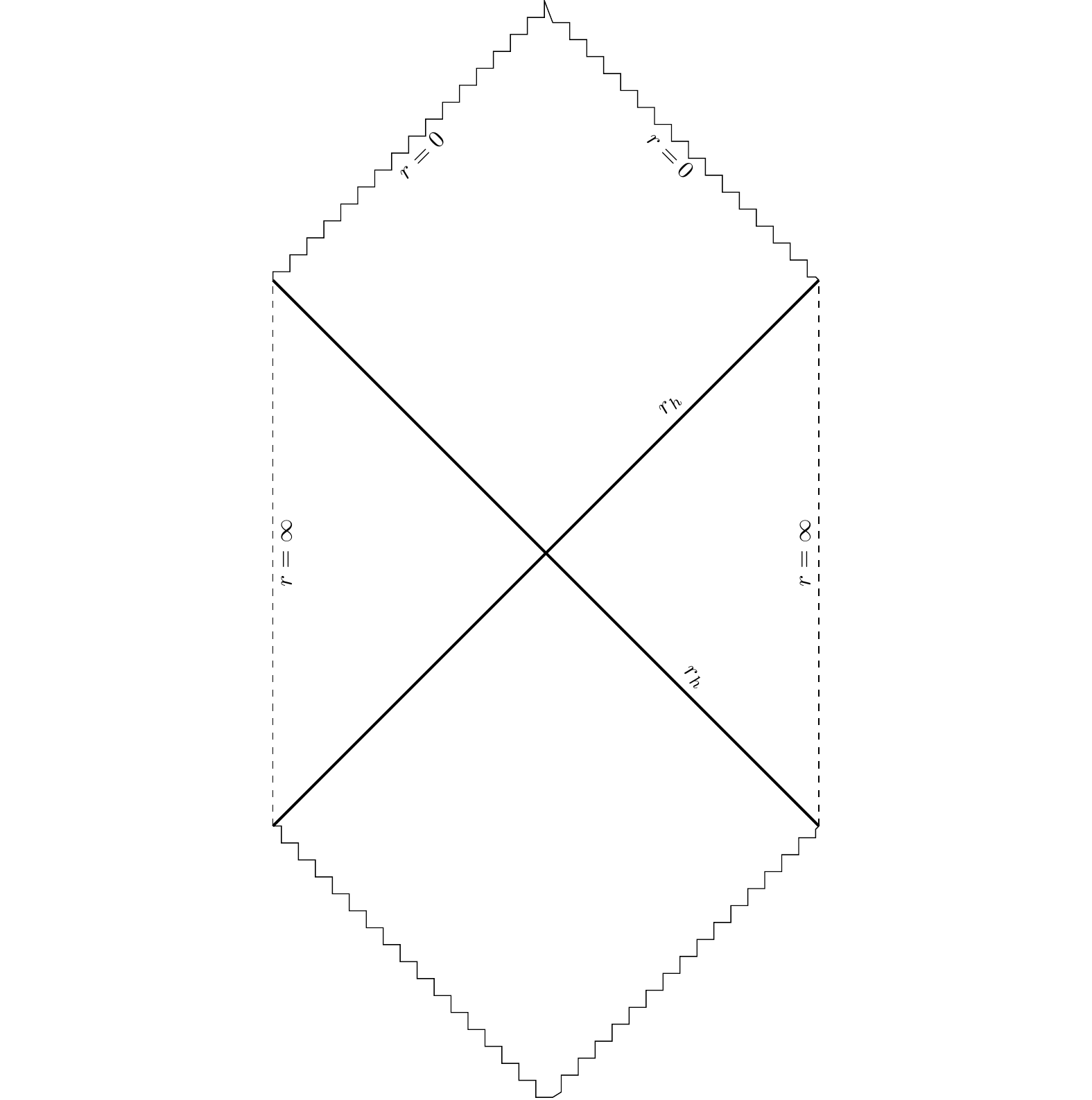} 
\caption{Qualitative Carter-Penrose diagram of the geometry Eq.(\ref{eq:g_ansatz}, \ref{eq:solf_BH}). The main regions along the holographic coordinate (boundary, horizon and  singularity) are understood from Table \ref{tab:reg_krusk}. It is important to observe from the given characterization that, in general, we have a divided causal structure compatible with the black hole interpretation of the solution.}
\label{fig:Penrose_HSVBH_nh}
\end{figure}

Despite the geometric resemblance in different aspects with AdS spacetime, both Schr\"odinger and Lifshitz  have unique features at a global level as succinctly pointed out in \cite{Duval:2008jg, Copsey:2010ya,Ayon-Beato:2011rts,Lei:2013apa}. In addition, the presence of the hyperscaling violating factor appends complexity so as to predict  a global nature based only on a simple inspection of the metric and curvature tensors. For instance, we argued in Sec. \ref{Sec:Basics} how for generic values of the exponents, there is an unavoidable singularity even for zero temperature configurations. Along this lines, we consider it pertinent to probe the global character of our \emph{a priori} dubbed black hole solution.
A resorted tool to envision the global properties of the configuration is to construct the Carter-Penrose diagram. While there is no framework  that grants the diagram the character of a formal or exhaustive characterization method, at the very  least, the existence of a divided causal structure is revealed \cite{Chrusciel:2012gz}.
We begin by rewriting the line element of the solution \eqref{eq:solf_BH} through the introduction of a tortoise coordinate
\begin{align} \label{eq:tortoirse}
    r^{*}=\int \frac{ dr}{br^{z+1} f(r)}=-\frac{1}{zbr^z}{}_{2}F_{1} \left( 1,\frac{z}{\beta};1+\frac{z}{\beta}, \left( \frac{r_{h}}{r} \right)^\beta \right),
\end{align}
where ${}_{2}F_{1}$ is the hypergeometric function and in terms of which, the metric reduces to
\begin{align}   \label{eq:g2Dtort}
    ds^2_{2D}=b^2 r^{-2\theta}\left[  r^{2z}f\left( -du^2+d{r^{*}}^2 \right) + r^2\left( dudv+d\vec{x}^2 \right)\right],
\end{align}
where it must be understood $r=r(r^*)$ and $f=f(r^*)$. First thing to notice is that, after diffeomorphism \eqref{eq:tortoirse}, the resulting line element \eqref{eq:g2Dtort} is no longer singular at the horizon $f=0$. Taking advantage of this fact, we can focus on 2-dimensional slices of constant $u$ and $\vec{x}$, where the analogue to Schwarzschild's radial geodesics live. In order to keep track of the different coordinate domains and perform a proper analysis of the relevant regions, we need more information than the general integral provided by
\eqref{eq:tortoirse}. We opt for two strategies to proceed from this point on. In this section, we consider a near-horizon approximation
\begin{align}
 r^{-(z+1)}f &= \kappa \left(r-r_h \right)+\dots, \nonumber \\
 \Rightarrow \,\, r^*&\approx \frac{\ln \left( |r-r_h| \right)}{\kappa} \,\, \text{where} \,\, \kappa:= \left. \left( r^{-(z+1)}f \right)'\right|_{r=r_h}.
\end{align}
Notice that the last line defines two different regions depending on the value of $r$ in comparison to the horizon location 
\begin{align}
     r^*\approx\begin{cases}
			\ln \left( r-r_h \right)/\kappa, & \text{if $r>r_h$}\\
           \ln \left( r_h-r \right)/\kappa, & \text{if $r<r_h$}
		 \end{cases}
\end{align}
It is important to keep in mind that, with the previous result, the blackening function takes the form
\begin{align} \label{eq:ftort}
    |f(r^*)|\approx e^{\kappa r^*}.
\end{align}
Following the standard approach, we move to null coordinates $(x^+=u+v,\, x^-=v-u)$ where the 2-dimensional metric reads 
\begin{align} \label{eq:g2Dur}
    ds^2_{2D}\approx-2 b^2r^{-2\theta+z-1} \left( \pm e^{\kappa(x^{-}-x^{+})}  \right) dx^+ dx^-.
\end{align}
The sign inside the parenthesis is determined if we are  measuring  outside ($+$) or inside ($-$) the horizon. We keep this convention hereupon. The next step is to define two new Kruskal-like coordinates
$(U=\pm exp(\kappa x^+),\, V=exp(-\kappa x^-) )$ which lead to 
\begin{align} \label{eq:g2Dur}
    ds^2_{2D}\approx\pm  \frac{2b^2}{\kappa^2}r^{-2\theta+z-1}   dU dV.
\end{align}
The last coordinates cause $f$ to not to appear explicitly anymore in the metric of the slices. Together with the form \eqref{eq:ftort} of $f$, yield a convenient representation of the division in the causal structure of the space. The relevant regions are catured in Tab. \ref{tab:reg_krusk}.  
\begin{table}[h]
\begin{tabular}{@{} *4l @{}}    \toprule
\emph{ Region} & $r$\emph{-coord}  & $r^*$\emph{-coord} & \emph{Kruskal coord}  \\ \midrule
Boundary & $r\rightarrow\infty$ &  $r^*=0$ & $UV=1$    \\ 
 Horizon & $r=r_h$  & $r^*\rightarrow -\infty$  & $UV=0$ \\ 
Singularity & $r= 0$  & $r^*=\rightarrow\infty$  & $UV=\infty $    \\
 \bottomrule
 \hline
\end{tabular}
 \caption{\label{tab:reg_krusk}  Kruskal coordinates representation of the relevant regions of the spacetime. }
\end{table} 
Lastly, in this generic description, one can represent the entire spacetime in a finite region through the coordinate compactification $(\tilde{U}=\arctan(U),\, \tilde{V}=\arctan(V))$. The resulting picture Fig. \ref{fig:Penrose_HSVBH_nh} makes manifest the presence of an event horizon that covers the  singularity and isolates it from the rest of the universe. 

\subsection{Exact diagram for a special case: $\beta=z$ }

\begin{figure}[H]\centering
  \includegraphics[width=0.4\textwidth]{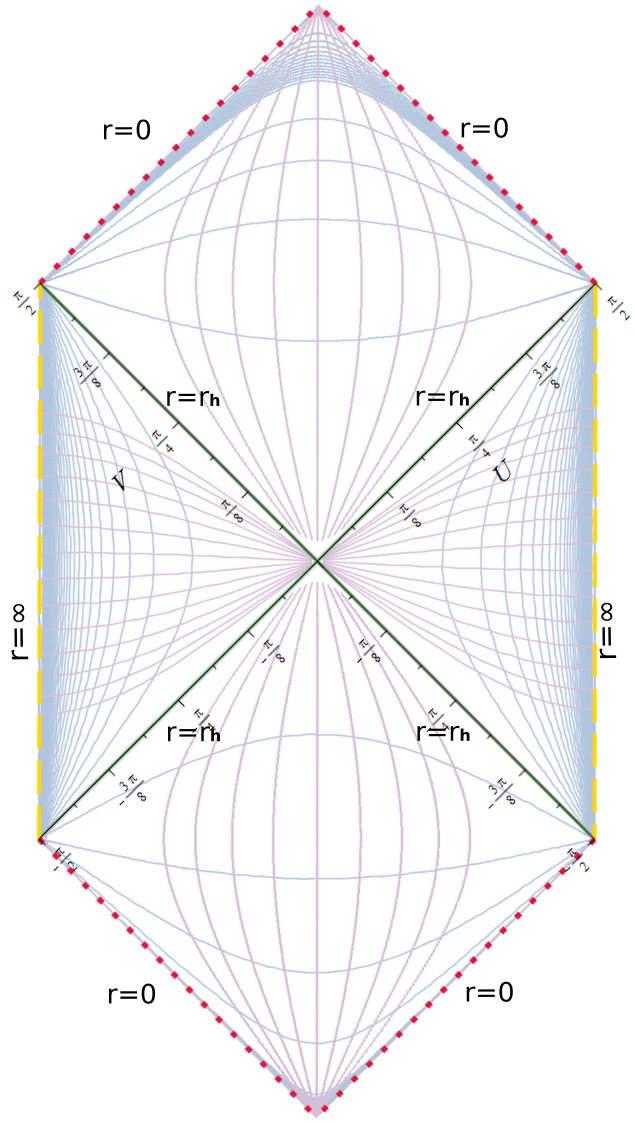} 
\caption{Carter-Penrose diagram of the black hole spacetime constructed in Sec. \ref{Sec:Solutions}. The central green cross represents the location of the event horizon dividing the exterior universe (left and right triangles) and the interior black hole (upper and lower diamonds). The boundary of the spacetime is located at the dotted yellow lines while the singularity is depicted by the dotted red lines. The spacial case $\beta=z$ allows us to analytically scrutinize the deformed light cone structure around the black hole. The lilac and blue lines are the constant $u$ and $r$ null geodesics respectively represented in the compactified space. }
\label{fig:Penrose_HSVBH}
\end{figure}

 Our second approach to the global characterization, is based on a fully analytically tractable case that depends on a particular parameter combination. 
Though we have already obtained qualitative information despite not having an explicit integral of the coordinate redefinition,  this approach is useful to provide a particular example compatible with the parameter bounds obtained so far. By setting $\beta=z $ (or equivalently, $\theta=(D-6+4z)/(D-2)$), the hypergeometric function \label{eq:tortoirse} dictating the tortoise coordinate reduces to a simple logarithm
\begin{align}
    r^{*}= \frac{ 1}{zb{r_h}^{z}} \ln\left(\left| 1-\left( \frac{r_{h}}{r} \right)^z \right|\right).
\end{align}
Equipped with it, we follow the same algorithm from the last section. The difference is that we will be able to provide the exact ranges of definition of the subsequent coordinates, even in regions away from the horizon. Having said this, we transform the 2-space into a manifestly lightcone form 
 \begin{align}   \label{eq:g2D+-}
    ds^2_{2D}=-2b^2 r^{2(z-\theta)} f dx^{+}dx^{-} \,\,\, \text{with }\,\, r=r\left(x^{+},x^{-} \right).
\end{align}
  The null coordinates take on the domain of $u$, which means that $-\infty< x^{\pm} <\infty$. The blackening function becomes $f=exp\left[  zb{r^z_{h}}(x^{-}-x^{+})/2 \right]$, and therefore the location of the horizon turns out to be  $x^{+}=\infty$ and $ x^{-}=-\infty$, just as in the Schwarzschild black hole. 
The $f$ term in front of \eqref{eq:g2D+-} can be reabsorbed with the introduction of the Kruskal-like coordinates,  
leading to a new form of the line element 
 \begin{align}   \label{eq:g2D+-2}
    ds^2_{2D}=-\frac{8}{z^2r^{2z}_h} r^{2(z-\theta)} dUdV \,\,\, \text{with }\,\, r=r\left(U,V\right).
\end{align}
 The metric function takes again a particularly simple form $f=\pm U V$, such that the horizon regions of relevance are again given as in Table \ref{tab:reg_krusk}. The compactification can be carried out in the same manner as 
$\left( U,V \right) \mapsto \left( \tilde{U}=\text{arctan}\left( U \right),\tilde{V}=\text{arctan}\left( V \right) \right) $ 
 such that the domain is given by $ -\pi/2\leq U \leq \pi/2$ and $ -\pi/2 \leq V \leq \pi/2$. Notice that while the horizon remains at $\tilde{U}=0$ or $\tilde{V}=0$, the metric is regular elsewhere except for the region $r(U,V)=0$ if $z-\theta<0$. The global structure of the spacetime is more easily described and understood based on the Carter-Penrose diagram Fig. \ref{fig:Penrose_HSVBH}.  Observe that the deformed light cones correspond to null geodesics of the $(u,r)$-plane brought to the compact $(\tilde{U},\tilde{V})$ representation. From it, we can learn that light rays coming from asymptotic infinity that cross the horizon have the singularity as the infinite future destination. Conversely, all geodesics in the interior region which are in causal contact with the singularity can not reach the exterior universe. Altogether, Fig. \ref{fig:Penrose_HSVBH} depicts the structure of a black hole spacetime with a proper horizon that ensures a cosmic censorship.  
  \vfill
 
\section{Thermodynamics \label{Sec:Thermo}}
 
  We devote the rest of this manuscript to capitalize on the physical significance of our solutions. As we have anticipated, the niche where we find a direct significance --- besides as an alternative gravity model--- is in the context of the holographic correspondence. In that regard, the systems at finite temperature 
 are in the spotlight since they have a prolific thermodynamics with appealing counterpart at the boundary. In the given lightcone-like coordinate $u,v$ can be interpreted as a bulk time and the other as the time at the boundary \cite{Herzog:2008wg}. However, we will prefer to switch to Schwarzschild-like coordinates
 \begin{equation}
 u=b\left(t+y\right),\qquad v=\dfrac{1}{2b}\left(y-t\right),   
 \end{equation}
where the metric acquires the structure
 \begin{align}
 ds^2=r^{-2\theta} & \left[-\left(b^4r^{2z}f+r^2\right)dt^2-2b^4r^{2z}fdtdy\right.\nonumber\\
 &\left.+\left(r^2-b^4r^{2z}f\right)dy^2+\dfrac{dr^2}{r^2f}+r^2d\vec{x}^2\right].
\end{align}
and the $t$-coordinate has a definite timelike norm outside the event horizon
\begin{equation}
 k=\partial_t\Longrightarrow k^2=-r^{-2\theta}\left(b^4r^{2z}f+r^2\right).
\end{equation}
and thus a unique interpretation of time.
 
 We start by working out the derivation of the total {mass}  by means of the quasi-local formalism introduced in \cite{Kim:2013zha}. In this method, the conserved charge associated to a Killing vector field $\xi$ is obtained through the integral
\begin{equation} \label{eq:conserved_charge}
\mathcal{Q} (\xi) = \int d^{D-2} x_{\mu \nu} \left(\Delta K^{\mu \nu}(\xi) -2 \xi^{[\mu}\int_{0}^{1} ds \Theta^{\nu]} (\xi|s) \right),
\end{equation}
where the antisymmetrized hypersurface integration element $d^{D-2}x_{\mu \nu}$ is defined through the Hodge dual ($\star$) of the $(D-2)$-volume form 
\begin{align*}
d^{D-2}x_{\mu \nu}= \frac{1}{2} \star\left( dx^{\mu_{1}} \wedge\ldots \wedge dx^{\mu_{D-2}}\right)_{\mu\nu},
\end{align*} 
whereas $s$ stands for a parameter allowing to interpolate the black hole configuration between the solution of interest ($s=1$) and the asymptotic one  ($s=0$). Furthermore, $\Delta K^{\mu \nu} (\xi)= K^{\mu \nu}_{s=1} (\xi)-K^{\mu \nu}_{s=0} (\xi)$ stands for the total difference of the Noether potentials between the two end points of the path, $s=1$ and $s=0$.  Finally, $\Theta^{\nu}$  is the surface term obtained after the variation of the corresponding action.
The Noether potential and the surface terms associated to our case are respectively given by 
\begin{align} \label{eq:NoetherP}
K^{\mu \nu}(\xi) & = 2 \sqrt{-g} \left[\frac{\nabla^{[\mu}\xi^{\nu]}}{2 \kappa}-\frac{1}{2}\frac{\partial \mathcal{L}}{\partial (\partial_{\mu} A_{\nu})}  \xi^{\sigma}A_{\sigma}\right], 
\\
\label{eq:surfaceT}
\Theta^{\mu}(\delta g, \delta \phi , \delta A ) & =  2 \sqrt{-g} \left(  \frac{g^{\alpha[ \mu }\nabla^{\beta]} \delta g_{\alpha \beta}}{2 \kappa}+\frac{1}{2} \frac{\partial \mathcal{L}}{\partial (\partial_{\mu} A_{\nu})} \delta A_{\nu}\right.\nonumber\\ 
& \left.+ \frac{1}{2} \frac{\partial \mathcal{L}}{\partial (\partial_{\mu} \phi)} \delta \phi\right).
\end{align}
The mass, as expected, is the conserved quantity consequence of the time translation along $\xi=-\partial_t$. For our results, the last expressions reduce to 
\begin{align}\label{eq:NoetherContrib}
 K^{\mu\nu}&=-2(\theta-1)r^{D-1-\theta(D-2)}\sqrt{\tilde{f}}\,\delta_t^{[\mu}\delta_r^{\nu]},
\\
\frac{\Theta^{\mu}}{r^{D-1-\theta(D-2)} } & =\left\{\dfrac{(\theta-1)}{\beta}\left[\dfrac{\beta-\left(\beta+2(D-2)\theta\right)\tilde{f}}{s\sqrt{ \tilde{f}}}\right]\right.\nonumber\\
& \left.-\sqrt{(\theta-1)(D-2)}\sqrt{\beta+\alpha \tilde{f}}\phi_0\right\}\delta_r^{\mu},  
\end{align}
with the help of the introduction of $\tilde{f}(r;s):=1-sM/r^\beta$, which is a deformation of the blackening function with the interpolation parameter $s$. 
Notice that we have redefined the arbitrary constant $r_h^\beta=M$ for further convenience in the present computation and that the interpolation was performed over all the independent integration constants $M$, $\phi_0$ and $q$. It is convenient to keep track of each contribution to the mass term, thus we show explicitly the indefinite integral version of the surface term,
\begin{widetext}
\begin{align}\label{eq:SimplecticContrib}
\int{\Theta^{\mu}ds}=r^{D-1-\theta(D-2)}\Bigg\{\dfrac{(\theta-1)}{\beta}& \left[-2(D-2)\theta\ln\left(\dfrac{1-\sqrt{ \tilde{f}     }}{1+\sqrt{ \tilde{f}}}\right)-2(\beta+2(D-2)\theta)\sqrt{ \tilde{f}}\right]\nonumber\\
&+\dfrac{2\sqrt{(\theta-1)(D-2)}}{3\alpha }\dfrac{\left(\beta+\alpha\tilde{f}\right)^{3/2}}{1{-}f}\phi_0\Bigg\}\delta_r^{\mu}, 
\end{align}
bearing in mind that this is the expression that has to be evaluated between the extreme points. Plugging in the previous results \eqref{eq:NoetherContrib} and \eqref{eq:SimplecticContrib} into the conserved charge \eqref{eq:conserved_charge} we get
\begin{align}\label{eq:GlobalCharge1}
\mathcal{M}=\lim_{r\rightarrow\infty}\Bigg[r^{D-1-\theta(D-2)}\left\{\dfrac{4(\theta-1)}{\beta}\right.& \left[(D-2)\theta\ln\left(\dfrac{1-\sqrt{\tilde{f}}}{1+\sqrt{\tilde{f}}}\right)+2(D-2)\theta\sqrt{\tilde{f}}\right]\nonumber\\
&\left.-\dfrac{4\sqrt{(\theta-1)(D-2)}}{3\alpha}\dfrac{\left(\beta+\alpha\tilde{f}\right)^{3/2}}{1-f}\phi_0\right\}V_{D-2}\Bigg]_{s=0}^{s=1}, 
\end{align}
\end{widetext}
where $V_{D-2}$ represents the Euclidean volume of the orthogonal sector to $t$ and $r$.  Since the overall integral is not quasilocal, we take the limit at spatial infinity were global charges should be evaluated. Also, it is important to remark that the only consistent way to get a non-zero but finite contribution with this method is to fix the hyperscaling violating parameter as
\begin{align*}
\theta=\dfrac{D-1}{D-2}.
\end{align*}
All in all we get the total mass
\begin{equation}\label{eq:Mass0}
 \mathcal{M}=2\sqrt{2\theta(\theta-1)(D-2)}\phi_0V_{D-2}.
\end{equation}
 The other global charge  is associated with the electromagnetic potential. It turns out that our configuration is indeed electrically charged according to
\begin{align}
 Q_e =\int{\star\,\chi(\phi)F} =qbV_{D-2}.
\end{align}
In the outcome, we spot that $q$ can be absorbed via a gauge election of $b$. This can be understood as if the electric charge is rigidly fixed, fact that plays an important role next in the thermodynamics.

Next, we compute the temperature though the standard Euclidean method. The Hawking temperature of our black hole solution can be computed from the Euclidean gravity by demanding that the
Euclidean black-hole solution should be smooth. As for
Lifshitz asymptotics \cite{Pedraza:2018eey,hartn}, one finds that the Euclidean time direction shrinks to a point in the event horizon. Smoothness at the horizon can be achieved by making the time direction periodic, this periodicity is directly identified with the inverse temperature of the black hole given that the time coordinate is identified as the one in dual field theory. Concretely we find
\begin{align} \label{eq:Temp}
 T  =\dfrac{b|f'(r_h)|}{4\pi}r_h^{z+1} = \dfrac{b\beta}{4\pi}r_h^z.
\end{align}
It is notorious that the hyperscaling violation parameter does not enter in the power of $r_h$, however, this result was already adverted in \cite{confinv}: a conformal factor should not appear in the general formula of the first line.
The entropy can be found according to Wald's prescription. In this case, there are no further contributions to that of the Bekenstein-Hawking area law. The reason being that the action does not contain higher corrections in the curvature
\begin{equation}\label{eq:Entropy}
S=\dfrac{\mathcal{A}}{4G_D}\equiv\dfrac{V_{D-2}}{4G_D}r_h^{-(\theta-1)(D-2)},  
\end{equation}
where $G_D=\kappa/8\pi$ is the D-dimensional gravitational constant.

Putting all pieces together, we proceed to check the consistency of the deduced global charges. The first check is the first Law of black holes thermodynamics
\begin{equation} \label{eq:1stLaw}
\delta{\cal{M}}=T \delta S, 
\end{equation}
which reveals in turn, that the  additive constant of the scalar field (ground state of the dual field theory) has a functional form in terms of the horizon radius 
\begin{equation}
 \phi_0=\dfrac{b\beta}{4\kappa}\sqrt{\dfrac{(\theta-1)(D-2)}{2\theta}}\dfrac{r_h^{-(\theta-1)(D-2)+z}}{(\theta-1)(D-2)-z}.
\end{equation}
such that the actual value of the mass \eqref{eq:Mass0} is
\begin{equation}\label{eq:Mass}
 \mathcal{M}=\dfrac{b\beta}{2\kappa}\dfrac{(\theta-1)(D-2)}{(\theta-1)(D-2)-z} r_h^{-(\theta-1)(D-2)+z}  V_{D-2}.
\end{equation}
The first law in Eq. \eqref{eq:1stLaw} does not display an explicit contribution from the electric part. Despite we have a net charge distinct to zero, its variation is limited by the gauge freedom in $b$ as we adverted before.

\subsection{Scalar charge and conjugate potential}

Following the concepts introduced by Gibbons et al. \cite{Gibbons:1996af} and extended in \cite{Astefanesei:2018vga}, we investigate the possibility of a scalar charge present in our gravitational configuration. As it was pointed out before, the dilaton field \eqref{eq:phiBH} has a non-divergent piece which, despite not coming from a Gaussian law, contributes at spatial infinity. For this analysis we restrict to the geometries preserving the standard hyperscaling $\theta=0$. As a result, the profile solution simplifies to
\begin{align} \label{eq:phiBHred}
\phi(\rho)=& -\sqrt{ \frac{  4   (D-2) }{\beta} } \text{arcsinh}\left(  \sqrt{   \rho^{\beta} -1 } \right)+\phi_0,
\end{align}
where $\beta=-\alpha$ and $\alpha>0$ ensures consistency of the solution outside the horizon.
According to the literature, the scalar charge $\omega$ can be read off as the coefficient for the leading term in a power series as
\begin{align}
    \phi(r)\Big|_{r\rightarrow\infty}=\phi_{\infty}+\frac{\omega}{r^\beta}+\mathcal{O}\left( r^{-2\beta} \right),
\end{align}
with $\phi_{\infty}$ the zeroth-order constant in the asymptotic expansion. The so called extended Thermodynamics, consider a new term originated due to an effective scalar charge $\omega\neq0$. In our setup, the said prescription would yield a complete first law given by
\begin{align}
    \delta M=T \delta S+\Phi_{E} \delta Q_{E}+\left( \frac{\partial M}{\partial \omega} \right) \delta \phi_{\infty}
\end{align}
where the partial derivative with respect to the scalar charge acquires a simple form in our case 
\begin{align}
    \left( \frac{\partial M}{\partial \omega} \right)=-\omega.
\end{align}
A short computation on \eqref{eq:phiBHred} reveals that the two important quantities correspond to
\begin{align}
\phi_{\infty}=-\sqrt{\beta(D-2)} \pi+\phi_0, \quad\quad \omega=\sqrt{\beta(D-2)} r_{h}^{\beta/2}.
\end{align}
In principle, this implies that there is no net contribution in the differential version of the first law since $\phi_\infty$ shows no explicit dependence on $r_h$. This can be enforced by fixing the value of the dilaton at the horizon $\phi_0$ according to specific boundary conditions. However, in the finite expression, the Smarr formula, there could exist a scalar part given by
\begin{align}
    -\omega\phi_\infty=\beta(D-2)\pi  \left( 1- \frac{\phi_0}{\sqrt{\beta(D-2)} \pi} \right)r_h^{\beta/2},
\end{align}
where again, it can be turned off with the help of the arbitrariness of $\phi_0$ and specific boundary conditions.

\section{The effect of the hyperscaling violation and dynamical exponents on the thermodynamics
\label{Sec:Scalarized}}
 
 In Sec. \ref{Sec:ScalarFree} we performed a rigorous analysis of the $\theta=1$ scenario which happens to be a peculiar point in the parameter space with various implications. Most remarkable is the fact that this choice turns off the scalar field, much alike as it was found in \cite{Herrera-Aguilar:2020iti} with the critical dynamical exponent at $z=1$. There, the asymptotically Lifshitz (hairy) black hole is smoothly connected in that limit with an asymptotically AdS scalar-free black hole. As a consequence, the emergence of the scalar field when moving from isotropy $z>1$ can be understood as a scalarization-like phenomenon--- given two additional conditions.
 In this work, we found that the general scalar-free solution is disconnected from the $\theta=1$ limit in our solution, despite appearing admissible in the integrated configuration. Hence, spontaneous scalarization or scalarization-like mechanisms cannot be analytically inspected here.
 
 Nonetheless, we will exploit how the global charges calculated in Sec. \ref{Sec:Thermo} are sensible to varying the dynamical and hyperscaling-violating parameters suggesting, at least qualitatively, that the unknown smoothly configuration constructed strictly without the scalar exhibit a rather distinct thermodynamics. In that sense, we will compute phase diagrams that show how state variables evolve at different combinations of $z$ and $\theta$.
 
\begin{figure}[H]\centering
  \includegraphics[width=0.4\textwidth]{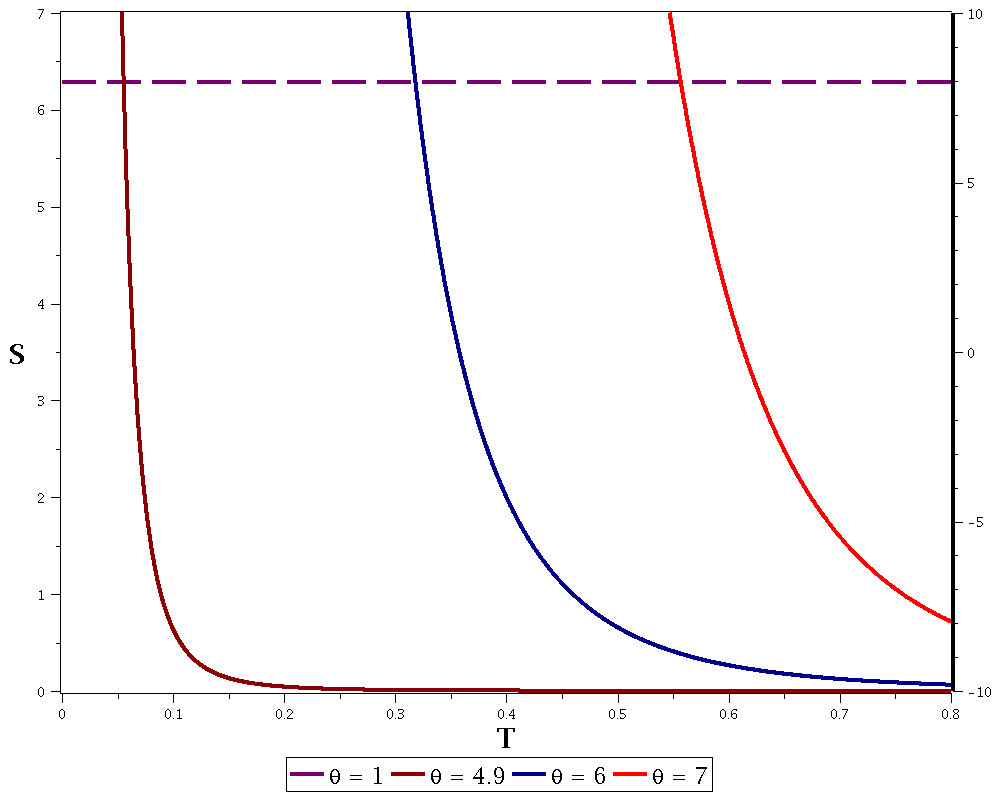} 
\caption{Entropy as function of the Temperature for the fixed reference values $D=5$, $z=3$, $V=1$, $\kappa=1$ and $b=1$.  We exemplify with families of curves at different $\theta>1$ and observe that the entropy for fixed temperature is always higher than that of the configurations at $\theta=1$.}
\label{fig:ST_beta_positive}
\end{figure}

\begin{figure}[H]\centering
  \includegraphics[width=0.4\textwidth]{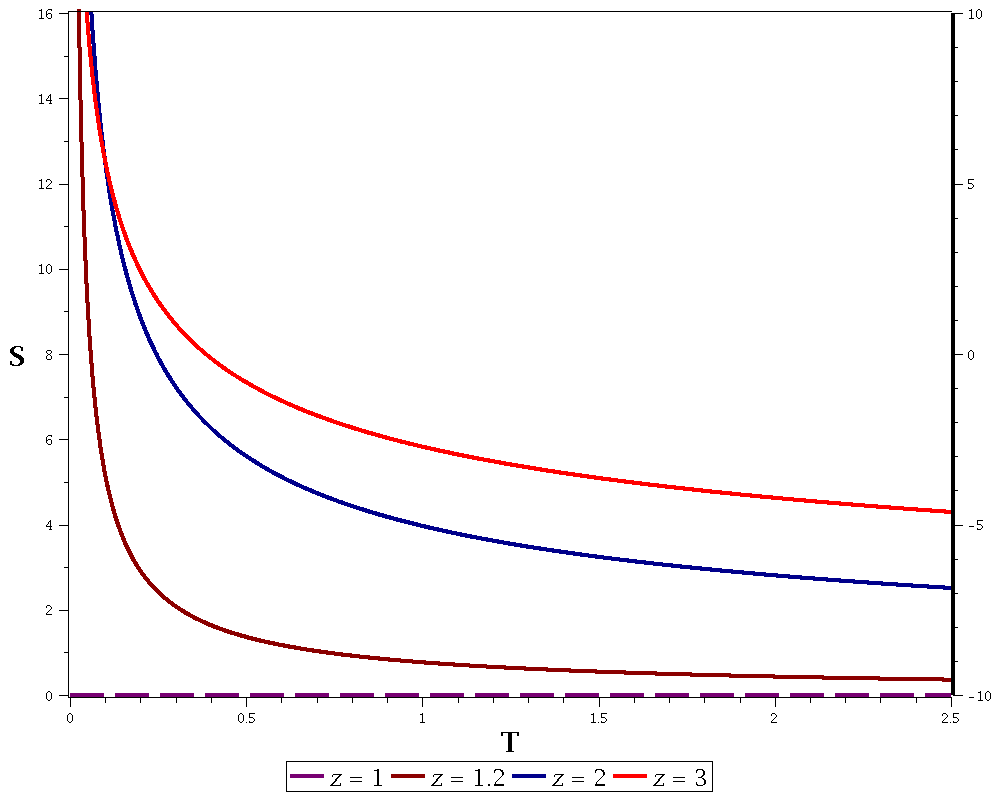} 
\caption{Entropy as function of the Temperature for the fixed values $D=5$, $V=1$, $\kappa=1$ and $b=1$. $T$ and $S$ are as in the restricted parameter caase \eqref{eq:TSM_simp}. It is displayed how configurations with $z>1$ away from isotropic scaling are favorable with higher entropy. }
\label{fig:ST_beta_negative}
\end{figure}

\begin{figure}[H]\centering
  \includegraphics[width=0.4\textwidth]{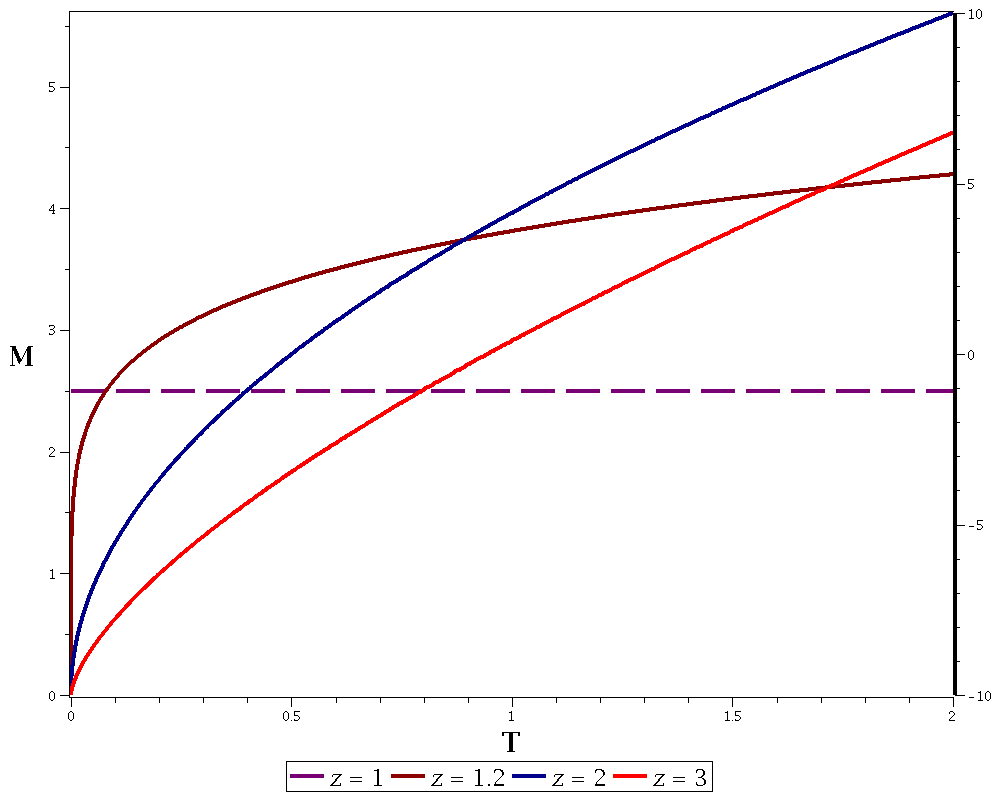} 
\caption{Mass as function of the Temperature for the fixed values $D=5$, $V=1$, $\kappa=1$ and $b=1$. $T$ and $M$ are as in the restricted parameter caase \eqref{eq:TSM_simp}. It is displayed how configurations with $z>1$ and sufficiently low temperature are preffered as a lower energy state.}
\label{fig:MT_beta_negative}
\end{figure}

To start our analysis, we look at the behaviour of the entropy \eqref{eq:Entropy} as function of the temperature \eqref{eq:Temp}. Since they both depend only on the horizon radius, one finds a power low relation between these quantities. To probe the effect of varying $\theta$ away from the scalar-free reference value. From Fig.~\ref{fig:ST_beta_positive}, we learn that generically, the configurations with  hyperscaling exponents $\theta>1$ are entropically prefered.
Next, we explore  the case of the mass \eqref{eq:Mass} ouput of the quasi-local formalism. Let us remind that in the tractable case, we have fixed  the hyperscaling violating exponent in terms of the dimension as $\theta=(D-1)/(D-2)$. Taking this into account, the relevant thermodynamical quantities acquire the simple structure
\begin{align} \label{eq:TSM_simp}
S & =\dfrac{2\pi}{\kappa}\dfrac{V_{D-2}}{r_h},\qquad T=\dfrac{b}{4\pi}5(z-1)r_h^z,\nonumber\\
M & =\dfrac{b}{2\kappa}5r_h^{z-1}V_{D-2}, 
\end{align}
which are independent from the dimension and then their functional form is only ruled by the critical dynamical exponent. We re-analyze the entropy-energy diagrams under said parameter constraints in Fig. \ref{fig:ST_beta_negative}
The entropies displayed in the last figure shows regions in the temperature where configurations with bigger degree of anisotropy are preferred but there are also exchange regions after which the contrary happens. However, similar to the previous scenario, the entropy for configurations with $z>1$ is always higher. This is confirmed by our last plot Fig.~\ref{fig:MT_beta_negative}. There probe now the diagrams of total energy  as a function of temperature. Again the relevant parameter is the critical dynamical exponent and thus we see the effect of displacing from isotropy. 
In the mass plot we see that there is a critical temperature under which the configutations with $z>1$ have a lower energy state and thus preferable with respect to the thermodynamics. Right at the critical temperature, both configurations with $z=0$  and $z>0$ would have the same energy. Finally, hotter black holes with $z>1$ would have higher energies than that of the isotropic case.

\section{Concluding remarks
\label{Sec:Remarks}}

We have constructed a novel family of anisotropic asymptotically Schr\"odinger black holes with arbitrary hyperscaling-violating and dynamical exponents within the framework of a generalized Einstein-Maxwell-dilaton model. Our starting point is different from the customary in the literature since we directly solve for the blackening function of the Schr\"odinger spacetime instead of starting from a Schwarzschild-like metric and modeling the Schr\"odinger behavior with the help of lightcone coordinate transformations \cite{Herzog:2008wg}. The advantage of this approach is that the hyperscaling-violating and critical dynamical exponents can be easily introduced but the computation of the global charges turn out be more involved. The black hole nature of our solution is explored by means of the Carter-Penrose diagram which reveals the global causal structure of the underlying Schr\"odinger geometry. We note that a similar global characterization of Schr\"odinger solutions have not been done before to our knowledge.

Notwithstanding, with the limitations of the quasi-local method we were able to construct a portion of the global charge thermodynamics compatible with the first law. Even more interestingly, we see that the particular value $\theta=1$ that regulates the trivialization of the dilaton field is also manifest in the behaviour of the energy and the entropy. Namely, it happens that for an arbitrarily wide region of the $T$-$S$ and $T$-$M$ diagrams, the configurations with a scalar field distinct to zero appear thermodynamically preferred. Similar to the scalarization phenomena, this suggests that the scalar field carries tachyonic modes over the hyperscaling-violating Schr\"odinger background. Thus, we spot a valuable followup work to determine the stability of this modes and compute the corresponding Breitenlohner-Freeman bound.

One interesting feature of the gravitational configurations here discussed arises from the Schr\"odinger spacetime \eqref{eq:gSchr}.  A simple computation shows that its associated Weyl tensor has a maximally degenerate principal null direction $\partial_v$, suggesting that at least in $D=4$ it belongs to the N type according to Petrov classification \cite{Coley:2004jv}. This is not surprising according to the arguments provided in \cite{Duval:2008jg,Ayon-Beato:2011rts}. In that sense, the zero temperature hyperscaling violating Schr\"odinger space can be understood as a conformal transformation of an AdS wave. However, despite  our \emph{ansatz} \eqref{eq:g_ansatz} can locally be also brought to this form for $r>r_h$, the diffeomorphism will be ill-defined for $r<r_h$ due to the change of sign of the blackening function. We have shown that there is indeed a causal division of the global structure. Consequently, the interpretation of the whole configuration is that of a black hole instead of gravitational radiation as suggested by the special algebraic type. There are other examples where  black hole solutions not necessary fall into a type D or type II classification, particularly when $D>4$. We refer the interested reader to \cite{Ortaggio:2010zg,Ortaggio:2014gma}.

Finally, is also remarkable that our black hole geometry supports different scalar field profiles as a consequence of the manifestly non-linear character of the differential system, a fact explicitly shown in appendix A.

\begin{acknowledgments}

JAMZ was partially funded by a grant from the Max-Planck-Society and the program ``ESTANCIAS POSDOCTORALES POR M\'EXICO 2021'', project 898686. JAMZ especially acknowledges the vast support provided by the MPI for Physics. The authors acknowledge financial support from a CONACYT Grant No. A1-S-38041. CERF acknowledges support from a CONACYT Grant No. 743421.
JAHM also acknowledges support from CONACYT through a PhD Grant No. 750974. DFHB is also grateful to CONACYT for a \emph{Postdoc por Mexico} Grant
No. 372516. All the authors are grateful to E. Ay\'on-Beato for enriching discussions. DFHB and JAMZ thank M. Bravo-Gaete for keen and useful observations.

\end{acknowledgments}

\appendix

\section{Non-unicity of the dilaton field}

A characteristic inherited from the form of the Schr\"odinger \emph{ansatz}  to the equations determining the  blackening function and the scalar field is that they happen to obey explicitly non-linear differential equations, unlike other spacetimes where the non-linearities do not actually appear. As we showed, these terms can be removed from the $f$ equation by means  of the coupling function. On the other hand, the non-linearity of the scalar field remains. In consequence, the unicity of the solutions, and  obtaining the most general solution for $\phi$ is not guaranteed. A similar problem was faced in \cite{Herrera-Aguilar:2020iti, Schr1}, where the ambiguity amounts to a sign change. 

 Our situation is different, and at least another independent solution for the dilaton in \eqref{eq:dilatonDEq} can be constructed. Despite it is not as physically relevant as the one previously presented, it is worth presenting for completeness and other possible applications. To clearly see this, let us write explicitly the dilaton's equation, preferably in terms of the $\rho$ variable for simplicity
 \begin{align} \label{eq:DEphi}
(\rho \phi')^2= & -\frac{ (\theta - 1)(D - 2)}{\rho^{2\theta-D(\theta-1)+5z-6}- 1} \Big\{ 2\theta \nonumber \\
&-\left[ D(\theta-1)-5z+6 \right]\rho^{2\theta-D(\theta-1)+5z-6} \Big\}.
\end{align}
The repeated power of $\rho$ and its factor in the numerator suggest a simplifying definition of constants. However, one might be tempted to factor out the $2\theta$ term in the numerator, in order to have a normalized form of equation. Namely, we can cast \eqref{eq:DEphi} in two seemingly equivalent forms
\begin{align} \label{eq:DEphiA}
(\rho \phi')^2= & \frac{ (\theta - 1)(D - 2)\left( \alpha \rho^\beta-2\theta \right)}{\rho^\beta-1},
\end{align}
and 
\begin{align} \label{eq:DEphiB}
(\rho \phi')^2= & \frac{ 2\theta(\theta - 1)(D - 2)\left( \hat{\alpha} \rho^\beta-1 \right)}{\rho^\beta-1},
\end{align}
which differ only in the constants 
$$
\alpha=D(\theta-1)-5z+6, \qquad\, \hat{\alpha}=\frac{D(\theta-1)-5z+6}{2\theta}
$$
and the overall factor $2\theta$. The first option is the path we followed in \ref{Sec:Solutions} with the already discussed scalar profile. The second option, integrates to a different parameter region and hence constitutes an independent solution, for instance, it is clear that $\theta=0$ is a singular point in contrast to \eqref{eq:DEphiA}.
Below we put forward the  configuration resulting from 

\begin{align}
\phi(\rho)&=\frac{\sqrt{2\theta(D-2)(\theta-1)}}{5z+D-6-\theta(D-2)}\Bigg[\ln\left(\frac{2-(1+\hat{\alpha})\rho^{\beta}+2K}{ \rho^{\beta}}\right)\nonumber\\&+\sqrt{\hat{\alpha}}\ln\left(1+\hat{\alpha}-2\hat{\alpha}\rho^{\beta}-2\sqrt{\hat{\alpha}}K\right)\Bigg],\label{eq:DEphiB}
\end{align}
where the following function was introduced
\begin{equation}
K(\rho)=\sqrt{\left(1-\alpha\rho^{\beta}\right)\left(1-\rho^{\beta}\right)}.
\end{equation}

 This field also satisfies all the equations of motion but describes a functionally independent scalar profile. We stress that, despite the properties of the inverse hyperbolic functions,  \eqref{eq:DEphiB} can not be mapped back to \eqref{eq:DEphiA}, they are intrinsically distinct. One must be careful with the parameter --- and spacetime--- regions where the field is real and finite. The main sensible difference with the alternative solution happens at the boundary $r\rightarrow\infty$ ($\rho\rightarrow 0$). In this solution, there is no parameter choice for which the field is regular. At the horizon $\rho=1$, the profile  drops off to zero. Inside the horizon, the reality will depend on a fine tuning between $z$ and $\theta$.

\section{A naked singularity}
\label{sing}
 In this appendix we present a geometry closely related to the black hole of Sec. \ref{Sec:Solutions}. Based on the same \emph{ansatz}, this solution is obtained through a less obvious choice of the non-minimal coupling function. Before, by setting $\chi$ as in \eqref{eq:chiBH0}, the nonlinear equation for the blackening function was reduced to a simple second order ODE.  Here, we pick the coupling in such a way that the nonlinear term still survives, while only the zeroth order term is cancelled out. Explicitly, 
 \begin{align}
 \chi(r)=\frac{q^2 r^{2[(D-3)(\theta-1)-z]}}{ 2b^2(z-1)[D-3+2z-\theta(D-2)]f^2 },\label{col}
 \end{align}
brings the $f$ equation to the form
\begin{align}
r^2f^{\prime\prime}+\frac{1}{2}{\frac{r^2f^{\prime2}}{f}}-\big [ \theta(D-2)-5(z-1)-D)\big]rf^\prime  =0.
\end{align}
Despite the non-linearity, it is not hard to show that the above has an equivalent algebraic equation; condition necessary and sufficient to determine all the independent solutions for the metric function. Namely, we find 
\begin{align}
f^{3/2}-c_{1}r^{2\theta -D(\theta-1)+5z-6}+c_{2} =0,
\end{align}
being $c_1$ and $c_2$ arbitrary integration constants. It is possible to construct up to three independent solutions in the complex plane. Nonetheless, the only purely real (and simpler) one is given by
\begin{align} \label{eq:NSsol}
f(r)=\left[  1 +\epsilon  \left( \frac{r_{h}}{r} \right)^{\beta} \right]^{\frac{2}{3}},
\end{align}
where $\epsilon= \pm 1$ is a free sign choice, and each selection will determine the nature of the resulting spacetime as we will detail soon. Notice that the exact same power $\beta$ pops-up in this solution too, despite being stemmed from a fundamentally disparate differential equation.  With $f$ as in  \eqref{eq:NSsol}, we proceed to integrate the scalar field from
\begin{align} \label{eq:SEq_NS}
\left( \phi_{\rho} \right)^2=\frac{ \theta(\theta-1)(D-2)\left( \gamma \rho^\beta -3\theta\right)}{ \rho^2 \left( \rho^\beta-1 \right) }.
\end{align}
We have used the variable $\rho$ previously introduced for economy. The constant $\gamma:=3\theta-\beta $ is analogous to $\alpha$ defined in Sec.~\ref{Sec:Solutions}.
Again, the result for the dilaton{s }profile is very similar to that of the black hole
\begin{align} \label{eq:f_NS}
\phi(\rho)=&  \sqrt{\frac{8(D-2) (\theta -1)}{3\beta^2}} \Bigg[ \sqrt{\gamma }\,\text{arcsinh}\left(\sqrt{ \frac{\gamma (1-\rho^\beta )}{\beta}}\right) \nonumber \\
& -\sqrt{3\theta } \,\text{arctanh}\left( \sqrt{ \frac{3\theta (1-\rho^\beta )}{3 \theta -\gamma \rho^\beta}}\right)\Bigg].
\end{align}
Following the same strategy as before, the self-interaction potential is obtained through evaluation of (\ref{eq:NSsol}) in (\ref{eq:PotAEq}).  It turns out as
\begin{equation}
V(\rho)=\frac{(D-2)(\theta-1)\left(\frac{r_h}{\rho}\right)^{2\theta}}{3\left( 1 +\epsilon \rho^{\beta}\right)^{1/3}}\left( \delta_5+ \delta_6\rho^{\beta}\right),\label{vagj1}
\end{equation}
with the following constants given by
 \begin{align*}
    \delta_5&=-3\left[\theta(D-2)-(D-1)\right], \\ \delta_6&=5z+2\theta(D-2)-(2D+3).
 \end{align*} 
  Finally, to determine the vector potential, we insert the metric function (\ref{eq:NSsol}) and the coupling function (\ref{col}) into the Maxwell's first integral (\ref{eq:Maxwell1st}) and fully integrate the vector field with respect to $ r $, obtaining the following solution
 \begin{align}
A(\rho)=\nonumber\frac{2b^2}{q}\left(\frac{r_h}{\rho}\right)^{D+2-\theta(D-2)-3z}\Bigg\{3(z-1)-\\\big[\theta(D-2)-(2z+D-3)\big]\rho^{\beta}\Bigg\}.\label{okl22}
\end{align}

The nature of this solution lies in the form  of the metric function $f$ (\ref{eq:NSsol}). Notice, that for the choice $\epsilon=1$, there is no real solution such that $f(r)=0$. As a consequence, it is not possible to form an event horizon for there is no null surface orthogonal to the asymptotically timelike Killing vector. In the other case, $\epsilon=-1$, despite being possible to solve $f(r)=0$ in the reals at $r=r_h$, it is easy to show that there is no signature change in the metric at that point. Even more deeply, the asymptotically Killing field has also no signature change in the whole range $0\leq r < \infty$, implying that an event horizon is not formed. In both scenarios the divergent point $r=0$ can not be removed as it is shown in the more general curvature invariants
   \begin{align}
   R=&{}\,(D-1)(\theta-1)r^{2\theta}\Big\{rf'\nonumber
   \\ &+\left[2-(\theta-1)(D-2)\right]f \Big\},
\label{invth1} \\
R_{\mu\nu} R ^ {\mu\nu}=&{}\,(D-1)(\theta-1)^2r^{4\theta}\Big\{\kappa_1f^2\nonumber
   \\ &+rf'\bigg[\kappa_3f+\frac{D}{4}rf'\bigg]\Big\},\label{ksm2}  \\
 R_{\alpha\beta\mu\nu} R ^{\alpha\beta\mu\nu}=&\, (\theta-1)^2r^{4\theta}\Big\{\kappa_2f^2\nonumber
   \\ &+(D-1)rf'\left[4f+rf'\right]\Big\},\label{ksm11}
\end{align}
with  $\kappa_{1,2}$ as in \eqref{eq:kconsts} and where a third constant appears 
\begin{align*}
  \kappa_3=& \,D-(D-2)(\theta-1).
\end{align*}

Unlike the black hole configuration, it is not possible to endow this geometry with a horizon, then the interpretation is that of a naked singularity.

\vfill


\end{document}